\documentclass[12pt]{article}
\usepackage{epsfig,hyperref}

\makeatletter
\def\vereq#1#2{\lower3pt\vbox{\baselineskip1.5pt \lineskip1.5pt
\ialign{$\m@th#1\hfill##\hfil$\crcr#2\crcr\sim\crcr}}}
\makeatother
\begin{document}

\begin{titlepage}
\begin{flushright}
{LBNL-46831\\ UCB-PTH-00/29\\ hep-ph/0009174\\September 2000}
\end{flushright}
\vskip 0.5cm
\begin{center}
{\large\bf Anarchy and Hierarchy\footnote{This work was supported in
part by the Department of Energy under contract DE--AC03--76SF00098,
in part by the National Science Foundation
under grant PHY-95-14797, in part by the Grant-in-Aid for Science
 Research, Ministry of Education, 
 Science and Culture, Japan
 (No. 12740146, No. 12014208).
}}
\vskip 0.5cm {\normalsize Naoyuki Haba}\\
\vskip 0.3cm {\it Faculty of Engineering, Mie University\\Tsu, Mie,
  514-8507, Japan} 
\vskip 1cm {\normalsize Hitoshi Murayama}\\
\vskip 0.3cm {\it Department of Physics, University of California \\
Berkeley, CA~~94720, USA\\ and \\ Theory Group, Lawrence Berkeley
National Laboratory\\ Berkeley, CA~~94720, USA}
\end{center}
\vskip .3cm
\begin{abstract}
	We advocate a new approach to study models of fermion masses and 
	mixings, namely anarchy proposed in Ref.~\cite{anarchy}.  In this 
	approach, we scan the $O(1)$ coefficients randomly.  We argue that 
	this is the correct approach when the fundamental theory is 
	sufficiently complicated.  Assuming there is no physical 
	distinction among three generations of neutrinos, the probability 
	distributions in MNS mixing angles can be predicted independent of 
	the choice of the measure.  This is because the mixing angles are 
	distributed according to the Haar measure of the Lie groups whose 
	elements diagonalize the mass matrices.  The near-maximal mixings, 
	as observed in the atmospheric neutrino data and as required in 
	the LMA solution to the solar neutrino problem, are highly 
	probable.  A small hierarchy between the $\Delta m^{2}$ for the 
	atmospheric and the solar neutrinos is obtained very easily; the 
	complex seesaw case gives a hierarchy of a factor of 20 as the most 
	probable one, even though this conclusion is more 
	measure-dependent.  $U_{e3}$ has to be just below the current 
	limit from the CHOOZ experiment.  The CP-violating parameter 
	$\sin \delta$ is preferred to be maximal.  We present a simple 
	$SU(5)$-like extension of anarchy to the charged-lepton and quark 
	sectors which works well phenomenologically.
\end{abstract}
\end{titlepage}
\setcounter{footnote}{0} 
\setcounter{page}{1} 
\setcounter{section}{0}
\setcounter{subsection}{0} 
\setcounter{subsubsection}{0}


\section{Introduction\label{sec:intro}}

The seemingly useless repetition of families has been a big puzzle in 
particle physics since the discovery of the muon.  Even more puzzling 
has been the fact that the particles in three families, despite their 
exactly the same gauge quantum numbers, have hierarchical masses and 
small mixing angles.  It appears, however, that we have lost the sense 
of the initial surprise having been accustomed to hierarchical masses 
and small mixings.  The very reason of the puzzle was the following 
simple naive expectation: the quantum mechanical states with the same 
quantum numbers are likely to have similar energies and should mix 
significantly.  Having tried to explain the hierarchy and small mixing 
angles for decades, we seem to have forgotten what the naive 
expectation was.

There was a revolution when the top quark was finally discovered.  We
were used to the idea that the weak gauge bosons are ``heavy'' while
the quarks, leptons are ``light.''  During the search for the top
quark for almost two decades, we used to question why the top quark is
so heavy, i.e., beyond the reach of the experiments.  After the
discovery of the top quark, however, we realized that the top quark
has the most natural mass among all quarks and leptons because its
Yukawa coupling to the Higgs boson condensate is $h_t = 1.0$ similarly
to the gauge coupling constants $e = 0.3$, $g = 0.8$, $g_s=1.2$.  Now
we think that the puzzle is not why the top quark is heavy, but rather
why all the other quarks and leptons are light.

In our opinion, the near-maximal mixing in the atmospheric neutrino 
data from the SuperKamiokande experiment is another revolution of the 
same kind as the case with the top quark.  It simply demands us to go 
back to the naive expectation: unless there are definite quantum 
numbers for three neutrino flavor states to be distinguished, they 
should fully mix and have comparable masses.  We are proposing this 
simple change in the perspective.  One may suspect, however, that this 
naive expectation would not lead to the apparent ``maximal mixing'' 
nor the hierarchy between two $\Delta m^2$ values required to explain 
atmospheric and solar neutrino data.  It was pointed out that a seesaw 
mass matrix without a particular structure should appear random from 
the low-energy point of view (anarchy), and a random matrix can well 
account for the observed pattern of neutrino mass and mixings 
\cite{anarchy}.

Therefore, it appears phenomenologically viable to consider all mass 
matrix elements for neutrinos within the seesaw mechanism for small 
neutrinos to be $O(1)$ without a particular structure: anarchy.  
However, one should ask if any of the results based on this idea would 
depend on the particular choice of the measure.  We show that the key 
requirement is that the measure should not depend on the choice of the 
basis with which the matrix elements are defined.  Once this 
requirement of basis-independence is made, distributions over the 
mixing angles are determined by the invariant Haar measure of the 
diagonalization matrices as we will see in 
Section~\ref{sec:neutrinos}.

We first present the concept of anarchy in Section~\ref{sec:anarchy}.  
We argue that the anarchy is an alternative approach to traditional 
model building which is applicable to wider range of theories.  In 
Section~\ref{sec:neutrinos}, we show how the distributions in the MNS 
mixing angles can be obtained from anarchy.  The assumed lack of fundamental 
distinction among three generations of neutrinos implies that the 
basis-independence of the measure.  Then the distributions in mixing 
angles are obtained solely by the Haar measure of the group which 
diagonalizes the mass matrices.  The predicted distributions are 
contrasted to the experimental data in 
Section~\ref{sec:phenomenology}.  Anarchy, however, raises another 
question: why the charged leptons, down quarks, and especially up 
quarks have hierarchical masses and small mixings.  We introduce a 
simple approximate $U(1)$ flavor symmetry which can answer this 
question in Section~\ref{sec:hierarchy}.  In 
Section~\ref{sec:conclusions}, we conclude.

\section{Anarchy\label{sec:anarchy}}

The idea of anarchy is simple: all coefficients of operators which do
not have particular reasons to be small, {\it i.e.}\/, allowed by
symmetries of the theory, are $O(1)$, and do not have particular
pattern, {\it i.e.}\/, appear random from the low-energy point of
view.  A fundamental theory should determine the $O(1)$ constants, but
if it is sufficiently complex, the $O(1)$ constants indeed likely
appear random.  Even though it is difficult to quantify the
``random''ness of the $O(1)$ constants, what should appear random are
the parameters in the Lagrangian, such as Yukawa matrix elements,
rather than the physical observables such as masses and mixing angles
which are defined only after diagonalization of matrices in the
Lagrangian.   Therefore, we take the point of view that the individual
elements of the Yukawa or mass matrices are distributed randomly.  

The idea of random $O(1)$ constants is used commonly in many different 
contexts.  For instance, in the chiral Lagrangian description of the 
low-energy hadron physics, one often faces the situation that one 
cannot predict the coefficient of the operators in the chiral 
Lagrangian from the corresponding operator written in terms of quarks 
and gluon fields.  Even though there should be in principle one-to-one 
correspondence between the Wilson coefficient in the QCD and the 
coefficient of chiral Lagrangian operators, it is beyond the scope of 
current techniques.  In this kind of situation, one often revokes the 
``naive dimensional analysis'' where one assumes an $O(1)$ coefficient 
unless there is a symmetry reason that the coefficient is suppressed.  
For instance, if two chiral Lagrangian operators contribute to a 
single process at the same order, people often argue that there should 
not be any severe cancellations between two contributions unless there 
is a reason for it.  Similar situation arises, for instance, if some 
of the particles in the Standard Model are actually composite.  From 
the point of view of the low-energy effective theory, the couplings 
are all $O(1)$ without a particular pattern: random.

In most compactifications of the superstring theory, Yukawa couplings 
among light degrees of freedom are $O(1)$, sometimes zero.  It is 
quite unusual to have small Yukawa couplings such as $10^{-6}$ for the 
electron whenever there is no reason that they are suppressed.  From 
the point of view of the low-energy effective field theory, the 
$O(1)$ couplings do not seem to have any particular pattern: random.

In grand-unified theories (GUTs) or Froggatt--Nielsen models of flavor, 
one introduces (quite a few) vector-like particles which become 
massive at a certain energy scale and decouple.  Even if all coupling 
constants are very close to unity, the low-energy coupling constants 
depend on many coupling constants, many VEVs and many inverse masses, 
and small deviations of coupling constants from unity get 
amplified.\footnote{In fact, the seesaw mechanism can be viewed as the 
simplest example, which does amplify the hierarchy, as seen in 
Section~\ref{subsec:spectrum}.} In the end we expect that the 
low-energy coupling constants are somewhat randomly distributed around 
unity.  Even the famous ``prediction'' of the $SU(5)$ 
grand-unification $m_{D} = m_{L}^{T}$, where $m_{D}$ and $m_{L}$ are 
down-quark and charged-lepton Dirac mass matrices, is likely to be 
spoiled by the mixing of many vector-like particles at the GUT-scale 
picking the GUT-breaking VEVs.  In the limit of large number of 
vector-like particles mixed with each other, the low-energy coupling 
constants become completely randomized.

In all of the above examples, seemingly random $O(1)$ coefficients 
arise in the low-energy effective theories as a consequence of the 
lack of precise knowledge on the fundamental theory.  Once the 
dynamics of the fundamental theory is completely understood, which we 
hope to do with the lattice QCD on the first example, the seemingly 
random $O(1)$ numbers can be predicted.  Still, in the absence of a 
fundamental theory behind the fermion masses and mixings, as opposed 
to the case of the hadronic physics where we know that the QCD is the 
fundamental theory, we should attempt to obtain some quantitative 
results without relying on the details of the dynamics.

We know of a perfectly sensible and beautiful theory of this kind: 
statistical mechanics.  When studying a collection of an astronomical 
number of particles, {\it e.g.}\/, Avogadro number, it is hopeless to 
measure the positions and the momenta of each particle precisely to 
predict the configuration of the particles in the future, even in the 
deterministic classical mechanics.  However the large number of 
particles allow us to believe that the motion of particles is 
completely randomized under various symmetry requirements, such as 
conservation of charges, number of particles, energy, {\it etc}\/, at 
least on average.  We can draw quantitative conclusions out of the 
randomness on the statistical basis.  We cannot predict the precise 
configuration of particles at a given instance.  However, we can 
predict the distributions in momenta, energies, positions, {\it etc}\/, 
with confidence.

What we advocate is a similar statistical treatment on the seemingly 
random $O(1)$ coefficients in low-energy effective theories consistent 
with certain symmetry requirements.  We scan over the $O(1)$ 
coefficients randomly.  However, an important question that arises 
here is what measure to use for the random scan.  In the case of 
statistical mechanics, Boltzmann distribution is justified as a 
consequence of the interaction between the relevant system with the 
infinitely large heat reservoir.  In our case, the choice of a 
particular distribution appears unwarranted.  Nonetheless, one may 
obtain distributions out of the random scan over the $O(1)$ parameters 
which do not depend on the choice of a particular measure; then such 
distributions can be regarded as rigorous predictions of the {\it 
anarchy}\/.\footnote{It is interesting to note that one of the 
definitions of the word ``anarchy'' is ``a utopian society of 
individuals who enjoy complete freedom without government'' according 
to Merriam-Webster's Collegiate Dictionary \cite{M-W}.  This 
definition is commensurate with what anarchism is: ``a political 
theory holding all forms of governmental authority to be unnecessary 
and undesirable and advocating a society based on voluntary 
cooperation and free association of individuals and groups'' 
\cite{M-W}.  Therefore we can hope to obtain phenomenologically 
successful theory of neutrino masses and mixings out of seemingly 
random numbers from anarchy.} As we will see below, the mixing 
matrices among neutrinos can be predicted from the anarchy and be 
compared to the experimental situations.

The anarchy is a complimentary approach to the traditional model 
building.  Traditionally, one attempts to write down a complete model 
of flavor by imposing certain symmetry which (hopefully) forbids all 
unwanted operators while explaining the observed structure of masses 
and mixings.  We would call this ``monarchy'' approach, as one aims 
for a model which is constrained tightly by the symmetries imposed and 
the particle content assumed with little freedom so that the model is 
predictive.  Then one hopes that the nature chose that particular 
model.  The advantage of this approach is that the model is predictive 
by design.  The disadvantage is that the model tends to become rather 
contrived and is not necessarily be believable.  The whole idea is 
based on the optimism that the high-energy physics is rather simple so 
that we can (eventually) test it directly from the low-energy 
observables.  The anarchy is the opposite approach.  We impose as 
little as possible.  Based on the statistical method, we try to figure 
out the tendency of the consequences from a given framework on the 
low-energy physics.  This approach is certainly not predictive, while 
it can be powerful enough so that it is applicable even in the 
situation where the high-energy physics is extremely complicated.  
This way, we can test if a certain simple idea is viable or not 
without going into details of particular models.

\section{Anarchy of Neutrinos\label{sec:neutrinos}}

In this section, we apply the anarchy approach to the neutrino mass 
matrices.  The fundamental assumption of our analysis is that there is 
no physical distinction among three generation of lepton doublets.  We 
would like to test this assumption by working out its consequences 
using the statistical method as discussed in the previous section.  
The key requirement is that the measure should not depend on the 
choice of the basis with which the matrix elements are defined because 
there is no physical distinction among three generations by 
assumption.  Once this requirement of basis-independence is made, 
distributions over the mixing angles are determined by the invariant 
Haar measure of the diagonalization matrices.

\subsection{A Simple Two-by-two Case}

It is instructive to study the consequence of a simple two-by-two
Majorana mass matrix for neutrinos assuming that each independent
matrix element is distributed randomly.  Since a Majorana mass matrix
is a symmetric matrix, there are only three parameters
\begin{equation}
  M_\nu = \left( \begin{array}{cc} M_{11} & M_{12}\\M_{12} & M_{22}
      \end{array} \right),
	  \label{eq:M2}
\end{equation}
and we assume the matrix elements to be real for simplicity of the
discussion.  One can rewrite this matrix in terms of physical
observables,
\begin{equation}
  M_\nu = m_{\rm average} + \frac{\Delta m}{2}
  \left( \begin{array}{cc} -\cos2\theta & \sin2\theta \\
      \sin2\theta & \cos2\theta \end{array}\right).
\end{equation}
Based on the assumption that the matrix elements $M_{11}$, $M_{12}$,
$M_{22}$ are distributed randomly, we can qualitatively understand how
the mixing angle is distributed.  By rewriting the volume element
$dM_{11} \wedge dM_{12} \wedge dM_{22}$ in terms of the observables,
we find
\begin{equation}
  dM_{11} \wedge dM_{12} \wedge dM_{22} = 
  \Delta m dm_{\rm average} \wedge d(\Delta m) \wedge d\theta,
\end{equation}
and the volume is {\it flat}\/ write respect to the mixing angle.  (We 
will omit the wedge symbol below.)  Then the distribution in $\sin^2 
2\theta$ is naturally peaked at zero and the maximal angles,
\begin{equation}
  d\theta = \frac{1}{4\cos2\theta\sin2\theta} d(\sin^2 2\theta).
\end{equation}
Note that the measure $d\theta$ over the angle is the invariant Haar 
measure of the $U(1)$ group.

In the realistic situation, we should consider a complex matrix.  As 
suggested by the phenomenological success of Kobayashi--Maskawa theory 
of CP violation, Nature appears to have chosen complex Yukawa 
matrices.  Then all the three independent elements of the matrix 
Eq.~(\ref{eq:M2}) are complex.  Correspondingly, the diagonalization 
matrix is unitary.  It can always be diagonalized by a unitarity 
matrix $U$,
\begin{equation}
	M_{\nu} = U \left( \begin{array}{cc} m_{1} & 0\\ 0 & m_{2}
      \end{array} \right) U^{T},
\end{equation}
where the unitarity matrix $U$ can be parameterized as
\begin{equation}
	U = e^{i\eta} \left( \begin{array}{cc} e^{i\omega} & 0\\ 0 & e^{-i\omega}
      \end{array} \right)
	  \left( \begin{array}{cc} \cos\theta & \sin\theta\\ 
	  -\sin\theta & \cos\theta \end{array} \right)
	  \left( \begin{array}{cc} e^{i\phi} & 0\\ 0 & e^{-i\phi}
      \end{array} \right) .
\end{equation}
We chose this parameterization such that the angles $\eta$ and 
$\omega$ are unphysical.  They can be absorbed by rephasing into the 
definition of the flavor and mass eigenstates.  The angle $\phi$ is a 
CP-violating phase, even though it does not appear in neutrino 
oscillation for this two-flavor case.  The measure for the matrix 
elements can then be rewritten as
\begin{equation}
	d^{2}M_{11} d^{2}M_{12} d^{2}M_{22}
	= (m_{2}^{2}-m_{1}^{2}) dm_{1}^{2} d m_{2}^{2} dU,
\end{equation}
where the measure over the angles is the invariant Haar measure over 
the $U(2)$ group
\begin{equation}
		dU = d(\sin^{2}\theta) d\eta d\omega d\phi
\end{equation}
up to a normalization constant.  Note that the distribution in 
$\theta$ is peaked even more strongly at the maximal mixing
\begin{equation}
	d(\sin^{2}\theta) = 2\sin\theta\cos\theta d\theta
	= \frac{1}{4\cos2\theta} d(\sin^2 2\theta) .
\end{equation}

We therefore conclude that a mass matrix without a particular 
structure would lead to a near-maximal mixing quite ``often.''  This 
observation should be contrasted to the wide-spread perception that a 
near-maximal mixing in the atmospheric neutrino data is very special.

\subsection{Why Haar Measure?}

In general, the measure over the angles are given solely in terms of 
the invariant Haar measure over the relevant group.  This is true as 
long as that the measure does not depend on the basis with which the 
matrix elements are defined.  The proof is very simple.  Since we 
discuss seemingly random mass matrix of neutrinos here, the measure 
with which we scan the parameter space should not depend on a 
particular choice of the basis of neutrino states.  This is the 
central assumption of {\it basis-independence}\/.  The measure discussed 
above for the two-flavor case, linear w.r.t. the individual mass 
matrix elements, satisfies this property, because the change of basis 
transforms the elements homogeneously as a unitarity rotation.  In 
fact, one can show that the distributions are determined by the 
invariant Haar measure over the relevant group from the requirement of 
the basis-independence.  This observation allows us to separate the 
measure over the mass matrix elements in terms of eigenvalues and 
group transformations.

Taking the complex Majorana case as an example, any mass matrix can be 
written as
\begin{equation}
  M = U D U^T,
\end{equation}
where $U \in U(N)$ and $D = \mbox{diag}(m_1, m_2, \cdots, m_N)$ is a
real diagonal matrix.  Because of the invariance of the measure $dM$
under the change of basis $M \rightarrow V M V^T$, the measure
$dM$ should contain the measure over the group $dU$ which is invariant
under the left translation $U \rightarrow V U$.  Since $U(N)$ is
a compact Lie group, a left-invariant measure is also right-invariant.
Therefore, $dU$ should be the invariant Haar measure over the group
$U(N)$.  Then the measure $dM$ can be written as
\begin{equation}
  dM = f(m_1, \cdots, m_N) \prod_{i=1}^N dm_i dU.
\end{equation}
Here the yet-undetermined function $f$ is symmetric under the 
interchange of eigenvalues.  In the appendix it is shown that the 
weight function $f$ should contain the factor 
$\prod_{i<j}(m_{i}^{2}-m_{j}^{2})\prod_{i}m_{i}$, and for the case of 
the linear measure, this exhausts the weight function $f$.  The only 
possible change of the measure is to introduce a weight function which 
depends on {\it invariants}\/, such as ${\rm Tr}M^{\dagger} M$, ${\rm 
det}M$.  However, such weight functions can depend only on the {\it 
eigenvalues}\/ of the mass matrix, consistent with the 
basis-independence.  Therefore the measure over the angles obtained 
from the Haar measure cannot be changed by the choice of the measure, 
and hence the distributions of the angle above are predictions of 
anarchy.  Exactly the same argument goes through for the Dirac and 
seesaw mass matrices.

Of course one should consider the diagonalization of the charged 
lepton mass matrix we well, since the MNS matrix is the mismatch 
between two diagonalization matrices, one of the neutrinos and the 
other of the charged leptons.  However, the translational invariance 
of the Haar measure guarantees that the diagonalization of the 
charged lepton mass matrix can be absorbed into the Haar measure and 
hence the Haar measure gives the distributions of the MNS angles.  
We consider a random scan over both the neutrino mass and the charged 
lepton mass matrices.  For a particular pick of the charged lepton 
mass matrix $M_{l} = U_{l} D_{l} U_{e}^{-1}$, we still scan over the 
neutrino mass matrix $M_{\nu} = U_{\nu} D_{\nu} U_{\nu}^T$, which we assume 
to be a complex Majorana matrix again for definiteness.  The measure 
is given in general by
\begin{equation}
	dM_{l} dM_{\nu} = f(m_{l_{i}}, U_{e}) g(m_{\nu_{i}}) 
	dm_{l_{i}} dm_{\nu_{i}}	dU_{l} dU_{e} dU_{\nu}.
\end{equation}
The 
basis-independence is imposed only on left-handed lepton doublets, and 
hence the measure may depend non-trivially on $U_{e}$.  However, 
$U_{e}$ is an unobservable matrix.  We can integrate over $U_{e}$ and 
obtain an effective weight factor
\begin{equation}
	\tilde{f}(m_{l_{i}}) = \int dU_{e} f(m_{l_{i}}, U_{e}).
\end{equation}
Now we can use $U_{MNS} = U_{l}^{-1} U_{\nu}$ instead of $U_{\nu}$, 
and the translational invariance of the Haar measure guarantees that 
$dU_{l} dU_{e} dU_{\nu} = dU_{l} dU_{e} dU_{MNS}$, where $U_{l}$ is 
now unobservable and its measure can be dropped.  The measure simplifies 
to\footnote{For the case of the linear measure, $f(m_{l_{i}}) = 
\prod_{i<j}(m_{l_{i}}^{2}-m_{l_{j}}^{2})^{2}\prod_{i}m_{l_{i}}$ and 
$g(m_{\nu_{i}}) = 
\prod_{i<j}(m_{\nu_{i}}^{2}-m_{\nu_{j}}^{2})\prod_{i}m_{\nu_{i}}$ as 
shown in the appendix.}
\begin{equation}
	dM_{l} dM_{\nu} = \tilde{f}(m_{l_{i}}) g(m_{\nu_{i}}) 
	dm_{l_{i}} dm_{\nu_{i}}	dU_{MNS}.
\end{equation}
Therefore, the distributions for the MNS mixing angles and the 
CP-violating phases are predicted in terms of the Haar measure again.
The only implicit assumption made in this argument is that the scan 
over the charged lepton and neutrino mass matrices are uncorrelated.  
We believe this is a natural consequence of typical flavor models as 
the mass matrices of the fields with different gauge quantum numbers 
are generated independently from different chain of the 
Froggatt--Nielsen fields, for example.  

Because of this, distributions in the mixing angles are direct 
consequences of the group structure and do not depend on the details 
of how one defines the measure.  On the other hand, the measure for 
the eigenvalues can depend on the details because one can modify the 
weight factor $f(m_{i})$ which is a symmetric function of all 
eigenvalues without affecting the basis-independence of the measure.

\subsection{Three-by-Three case}

For three generation case, details are discussed in 
Appendix~\ref{sec:measures}.  Here we merely summarize the results.  
We decompose the linear measure over the real or complex elements of 
both Majorana and Dirac mass matrices into the measure over the angles 
and the mass eigenvalues.

For the real Majorana case, the symmetric real mass matrix is written 
as $M = O D O^T$ where $O$ is an $SO(3)$ diagonalization matrix and 
$D = {\rm diag}(m_{1},m_{2},m_{3})$.  Then we find
\begin{equation}
  dM = (m_{1}-m_{2}) (m_{2}-m_{3}) (m_{3}-m_{1}) dm_{1} dm_{2} dm_{3} 
  dO,
  \label{eq:majorana}
\end{equation}
where $dO$ is the invariant Haar measure of the $SO(3)$ group.  

For the real Dirac case, the mass matrix is written as $M = O_{L} D 
O_{R}^{T}$ where $O_{L,R}$ are $SO(3)$ diagonalization matrices and $D 
= {\rm diag}(m_{1},m_{2},m_{3})$.  Then we find
\begin{equation}
  dM = (m_{1}^{2}-m_{2}^{2})(m_{2}^{2}-m_{3}^{2})(m_{3}^{2}-m_{1}^{2}) 
  dm_{1} dm_{2} dm_{3} dO_L dO_R.
  \label{eq:dirac}
\end{equation}
Again $dO_{L}$, $dO_{R}$ are invariant Haar measures for the 
corresponding $SO(3)$ groups.

For the complex Majorana case, the mass matrix is written as $M = U D 
U^{T}$ where $U$ is a $U(3)$ diagonalization matrix and $D = {\rm 
diag}(m_{1},m_{2},m_{3})$.  Then we find
\begin{equation}
  dM = (m_{1}^{2}-m_{2}^{2})(m_{2}^{2}-m_{3}^{2})(m_{3}^{2}-m_{1}^{2}) 
  m_{1} m_{2} m_{3} dm_{1} dm_{2} dm_{3} dU,
  \label{eq:majoranac}
\end{equation}
where $dU$ is the invariant Haar measure of the $U(3)$ group.

For the complex Dirac case, the mass matrix is written as $M = U_L D 
U_R^\dagger$ where $U_{L,R}$ are $U(3)$ diagonalization matrices and 
$D = {\rm diag}(m_{1},m_{2},m_{3})$.  Then we find
\begin{equation}
  dM = 
  (m_{1}^{2}-m_{2}^{2})^{2}(m_{2}^{2}-m_{3}^{2})^{2}(m_{3}^{2}-m_{1}^{2})^{2}
  m_{1} m_{2} m_{3} dm_{1} dm_{2} dm_{3} 
  \frac{dU_L dU_R}{d\varphi_{1}d\varphi_{2}d\varphi_{3}}.
  \label{eq:diracc}
\end{equation}
Because of the reparameterization invariance $U_{L} \rightarrow U_{L} 
\Phi$, $U_{R} \rightarrow U_{R} \Phi$, with $\Phi = {\rm 
diag}(e^{i\varphi_{1}}, e^{i\varphi_{2}}, e^{i\varphi_{3}})$, the Haar 
measure $dU_{L} dU_{R}$ is modded out by this invariance.

Finally, we consider also seesaw neutrino mass matrices, which are 
arguably the most promising ones in order to explain the smallness of 
neutrino masses as well as the most motivated one from the point of 
view of grand unification.  In this case, we have Majorana-type 
right-handed neutrino mass matrix $M_{R}$ as well as Dirac-type mass 
matrix between left- and right-handed neutrinos $M_{D}$.  The full 
mass matrix is given by
\begin{equation}
	\left( \begin{array}{cc}
		0 & M_{D} \\
		M_{D}^{T} & M_{R}
	\end{array}
	\right),
\end{equation}
such that the full Majorana-type mass matrix is symmetric.  It is 
assumed that there is a large hierarchy between the eigenvalues of 
$M_{D}$ to those of $M_{R}$.  The mass matrix among the heavy (dominantly 
right-handed) neutrino states is just $M_{R}$ up to corrections of 
the order of $M_{D}^{2}/M_{R}^{2}$, while that among the light 
(dominantly left-handed) neutrino states is given by
\begin{equation}
	M_{D} M_{R}^{-1} M_{D}^{T},
\end{equation}
again up to corrections of the order of $M_{D}^{2}/M_{R}^{2}$.  This 
is the famous seesaw formula.  Hereafter we discuss only the complex 
case.  By diagonalizing both $M_{D} = U_{L} D_{D} U_{R}^{-1}$ with 
$D_{D} = {\rm diag}(m_{1},m_{2},m_{3})$ and $M_{R} = U_{M} D_{M} 
U_{M}^{T}$ with $D_{M} = {\rm diag}(M_{1},M_{2},M_{3})$, we find the 
linear measure to be
\begin{eqnarray}
	\lefteqn{dM_{D} dM_{R} =} \nonumber \\
	& &
	(m_{1}^{2}-m_{2}^{2})^{2}(m_{2}^{2}-m_{3}^{2})^{2}(m_{3}^{2}-m_{1}^{2})^{2}
  	m_{1} m_{2} m_{3} dm_{1} dm_{2} dm_{3} 
	\frac{dU_L dU_R}{d\varphi_{1}d\varphi_{2}d\varphi_{3}}
	\nonumber \\
	& &
	(M_{1}^{2}-M_{2}^{2})(M_{2}^{2}-M_{3}^{2})(M_{3}^{2}-M_{1}^{2}) 
	M_{1} M_{2} M_{3} dM_{1} dM_{2} dM_{3} dU_{M}.
	\label{eq:seesawc}
\end{eqnarray}
However, the light neutrino mass matrix is then given by $U_{L} D_{D} 
U_{R}^{-1} U_{M} D_{M}^{-1} U_{M}^{T} U_{R}^{-1T} D_{D} U_{L}^{T}$, 
and hence $U_{R}$ and $U_{M}$ are not separately physical but only the 
combination $U_{\rm rel}=U_{R}^{-1} U_{M}$.  Therefore one can 
simplify the measure $dU_{R} dU_{M}$ to just $dU_{\rm rel}$.  Note 
that one still needs to diagonalize the matrix
\begin{equation}
	\left( \begin{array}{ccc}
		m_{1} & &\\ & m_{2} &\\ & &m_{3}
	\end{array} \right) U_{rel}
	\left( \begin{array}{ccc}
		M_{1}^{-1} & &\\ & M_{2}^{-1} &\\ & &M_{3}^{-1}
	\end{array} \right) U_{rel}^{-1T}
	\left( \begin{array}{ccc}
		m_{1} & &\\ & m_{2} &\\ & &m_{3}
	\end{array} \right)
	\label{eq:seesawm}
\end{equation}
to find the mass eigenvalues of the light neutrinos.  On the other 
hand, the measure $dU_{L}$ can be replaced by $dU_{MNS}$ as discussed 
earlier. 

In each of the cases, we find that the distributions in the MNS mixing 
angles are determined solely by the Haar measure.

\subsection{Distributions in Angles}

Now that we have learned that the distributions in the MNS mixing 
angles are distributed according to the Haar measure, we would like to 
see how they look.  The distributions are different for the $SO(3)$ 
case (real mass matrices) and the $U(3)$ case (complex mass 
matrices).  

\begin{figure}
	\centerline{
	\psfig{file=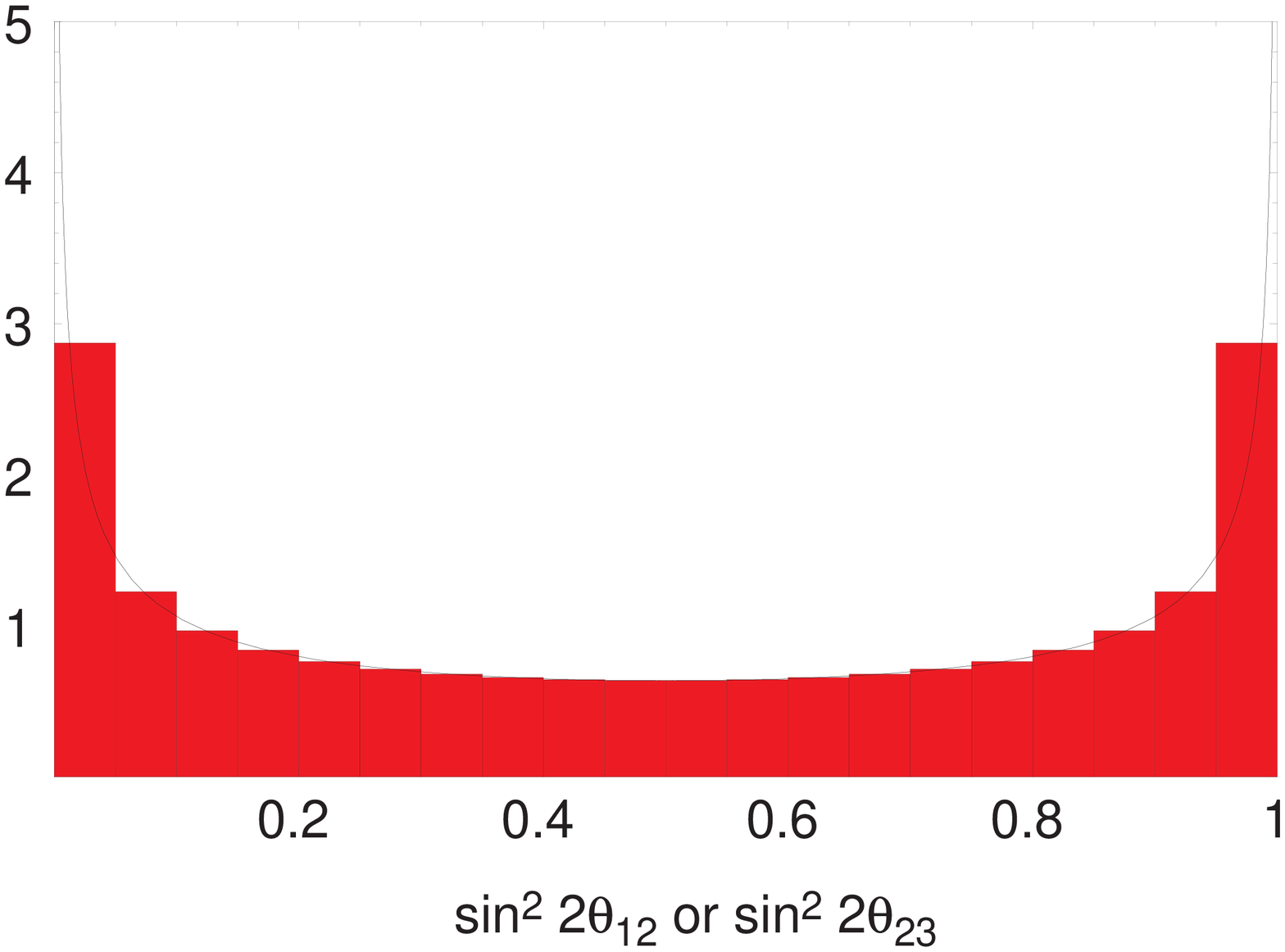,width=0.4\textwidth} \hfill
	\psfig{file=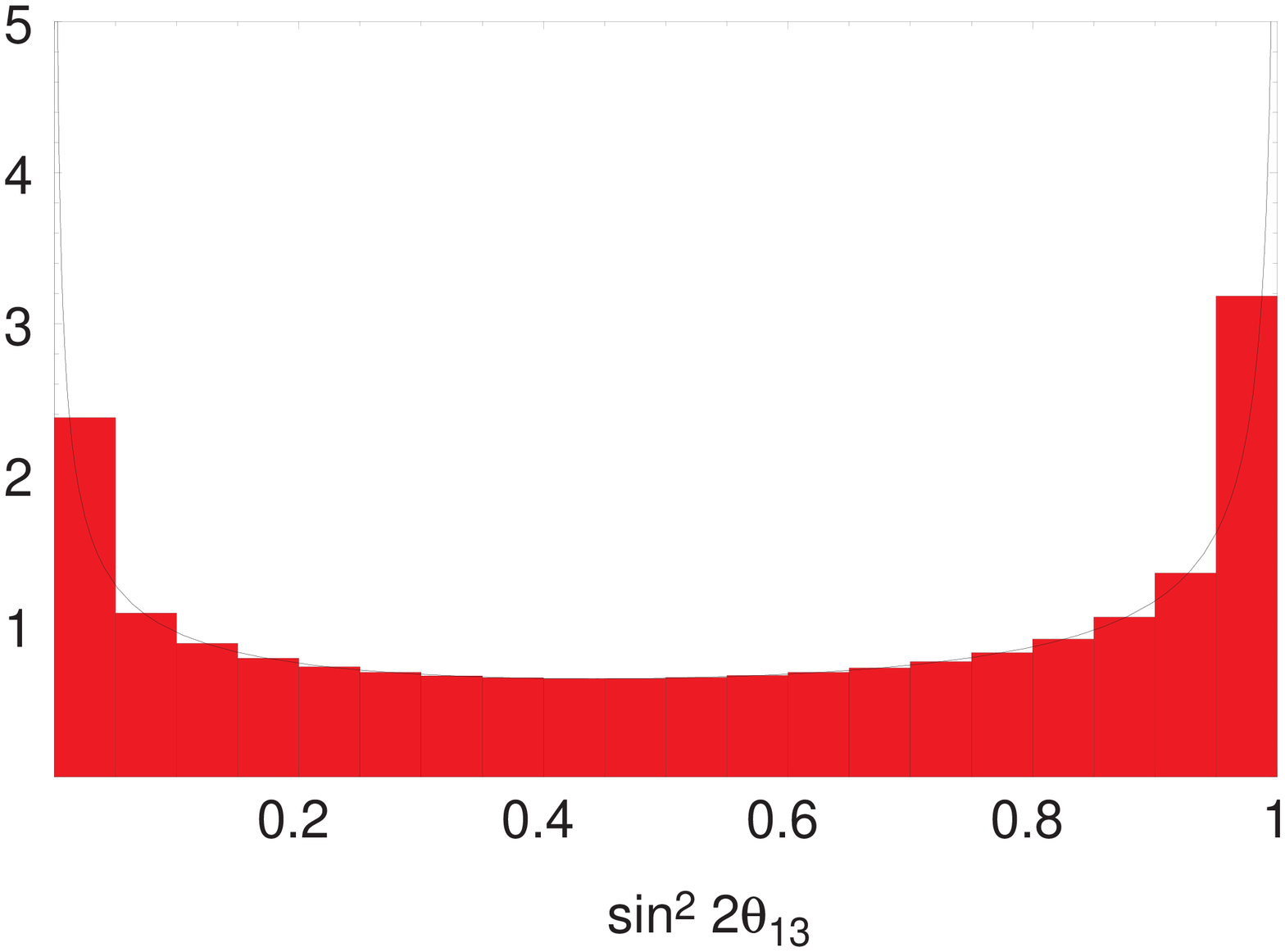,width=0.4\textwidth}
	}
	\caption{Distributions in (a) $\sin^{2} 2\theta_{12}$ or $\sin^{2} 
	2\theta_{13}$ and (b) $\sin^{2} 2\theta_{13}$ for the case of real 
	mass matrices.}
	\label{fig:SO3}
\end{figure}

First for the case of real matrices, the distributions in the MNS 
mixing angles are given in terms of the $SO(3)$ Haar measure (see 
Appendix~\ref{subsec:Haar})
\begin{equation}
  dO = \cos\theta_{13} d\theta_{12} d\theta_{13} d\theta_{23}.
\end{equation}
It is straightforward to plot the distributions against $\sin^{2} 
2\theta_{ij}$.  In Fig.~\ref{fig:SO3}, we show distributions in 
(a) $\sin^{2} 2\theta_{12}$ or $\sin^{2} 2\theta_{23}$, and (b) 
$\sin^{2} 2\theta_{13}$.  The solid lines show the differential 
distributions and the histograms the integrated probabilities with the 
bin size of $0.05$.

\begin{figure}
	\centerline{
	\psfig{file=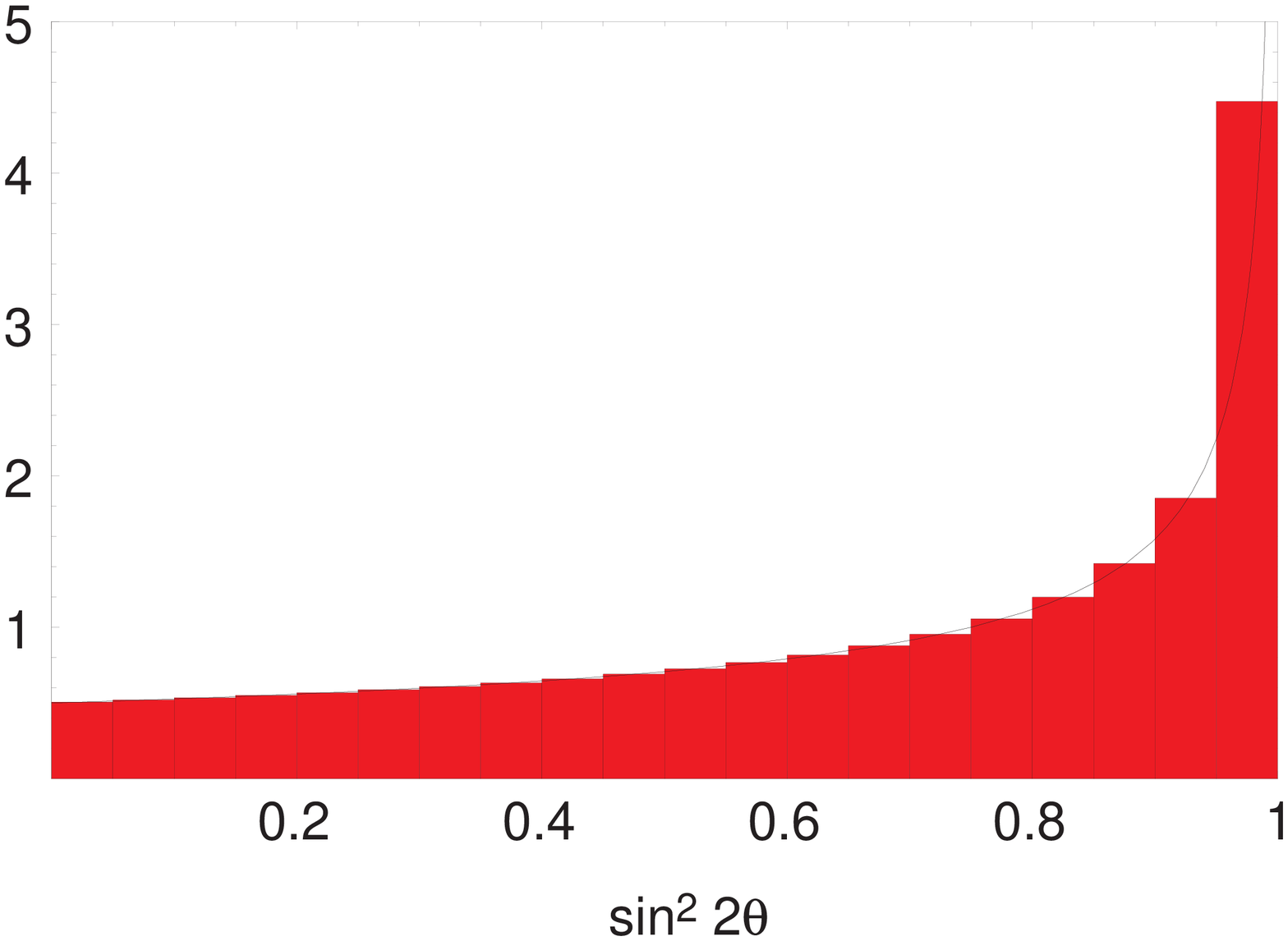,width=0.4\textwidth} \hfill
	\psfig{file=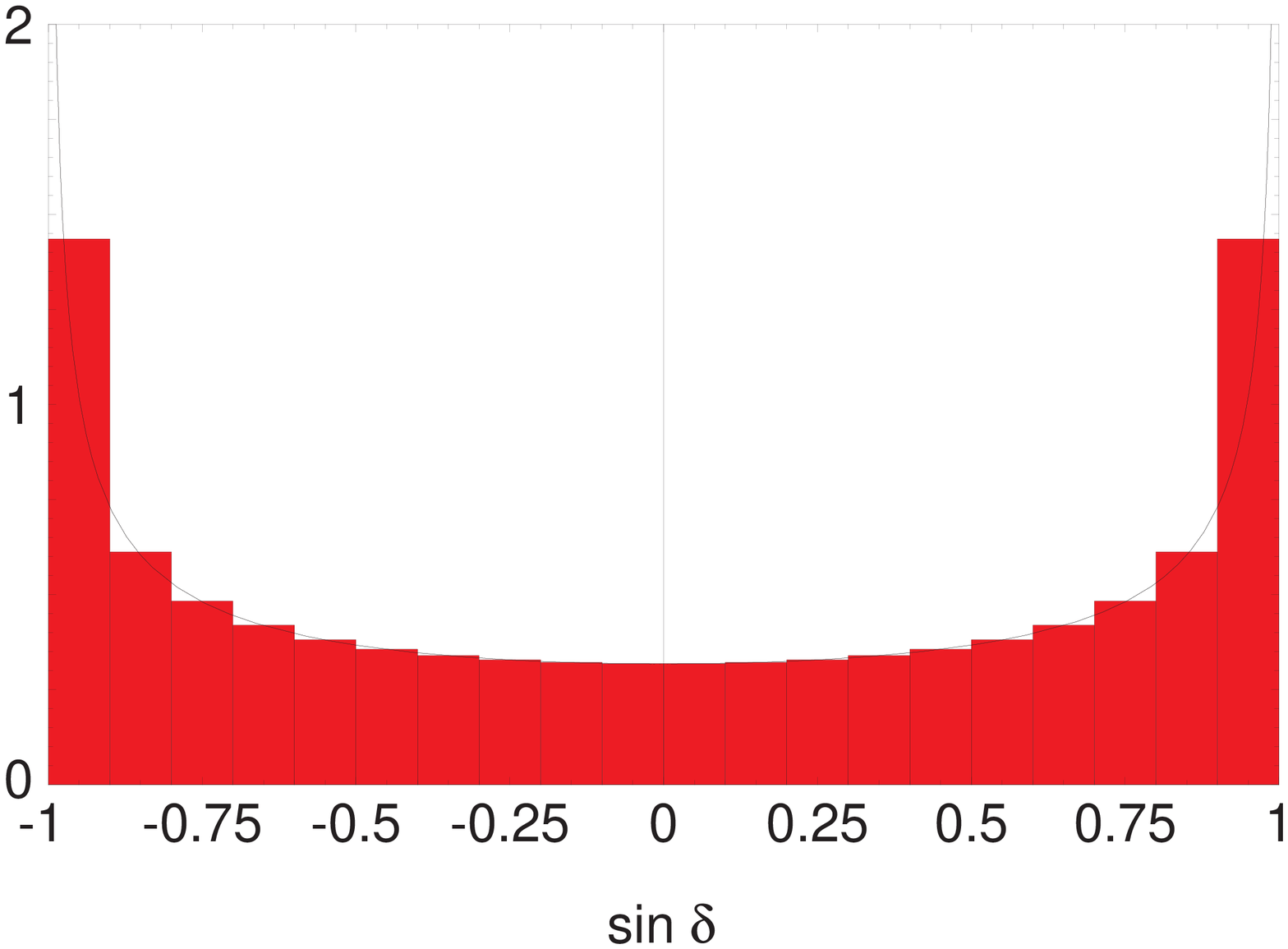,width=0.4\textwidth}
	}
	\caption{Distributions in (a) $\sin^{2} 2\theta_{12}$ or $\sin^{2} 
	2\theta_{13}$ or $\sin^{2} 2\theta_{23}$ and (b) $\sin\delta$ for 
	the case of complex mass matrices.}
	\label{fig:U3}
\end{figure}

Second for the realistic case of complex matrices, the distributions 
in the MNS mixing angles are given in terms of the $U(3)$ Haar measure 
(see Appendix~\ref{subsec:Haar})
\begin{equation}
	d U = d s_{12}^{2} d c_{13}^{4} d s_{23}^{2} d\delta
	d\eta d\phi_{1} d\phi_{2} d\chi_{1} d\chi_{2} .
\end{equation}
Since the angles $\eta$, $\phi_{1,2}$ are unphysical while the angles 
$\chi_{1,2}$ are Majorana phases relevant only to lepton-number 
violating processes, we focus only on $\sin^{2} 2\theta_{ij}$ as well 
as the CP-violating parameter $\sin \delta$ which is the dependence on 
$\delta$ in neutrino oscillations.  It turns out that the distributions 
in $\sin^{2} 2\theta_{ij}$ are all the same.  In Fig.~\ref{fig:U3}, 
we show distributions in $\sin^{2} 2\theta_{ij}$ and $\sin \delta$.  
The solid lines show the differential distributions and the histograms 
are integrated probabilities with the bin size of $0.1$.

The distributions shown in Ref.~\cite{anarchy} used different 
parameterization of the angles and are consistent with what we show 
for the case of real matrices except for the following small 
correction.  There, the elements of the mass matrices were scanned 
from $-1$ to $1$ independently, which forms a hypercube.  But such a 
scan unfortunately breaks the basis-independence as the orientation of 
the hypercube changes as the basis rotates.  This resulted in a small 
distortion of the distributions, which made them different for Dirac, 
Majorana, and seesaw cases.  We redid the scan with the same programs 
used in Ref.~\cite{anarchy} with a basis-independent boundary ${\rm 
Tr}M^{\dagger} M \leq 1$, which forms a hypersphere instead, and 
verified that the distributions obtained from the analytic arguments 
result from numerical scans.

\subsection{Typical Mass Spectrum\label{subsec:spectrum}}

The mass spectrum from random mass matrices depends in general on the 
particular choice of the measure, unlike the distributions in the 
angles.  This is because one can introduce an additional weight 
factor which depends only on the mass eigenvalues on top of the linear 
measure as discussed earlier.  In this subsection, we use the linear 
measure throughout, and show what mass spectrum results from this 
choice.  In order to keep basis-independence of the measure, we use 
the scanning region ${\rm Tr}M^{\dagger} M \leq 1$.  This translates 
to the condition $m_{1}^{2} + m_{2}^{2} + m_{3}^{2} \leq 1$, {\it 
i.e.}\/, to the region inside a three-dimensional sphere. 

\begin{figure}
	\centerline{
	\psfig{file=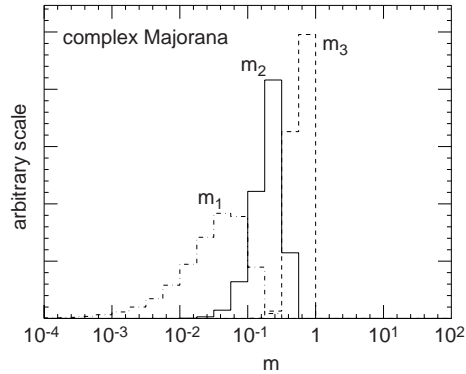,width=0.4\textwidth}
	}
	\caption{The distributions of mass eigenvalues for the complex 
	Majorana case with the linear measure.}
	\label{fig:mlog10m}
\end{figure}

The first distribution we show is the mass spectrum in the complex 
Majorana case, with the measure Eq.~(\ref{eq:majoranac}).
The distributions in Dirac and real cases are similar and we focus on 
this case.  We take the convention $0<m_{1}<m_{2}<m_{3}$ without a 
loss of generality.  Fig.~\ref{fig:mlog10m}, we show distributions 
in three mass eigenvalues.  We can see that there is a sharp cutoff at 
$m_{3} = 1$ because of the condition $m_{1}^{2} + m_{2}^{2} + 
m_{3}^{2} \leq 1$.  An interesting point is that the other eigenvalues 
are often much smaller than $O(1)$, almost a half (one) order of 
magnitude for $m_{2}$ ($m_{1}$).  The mass spectrum is somewhat 
hierarchical.  This is rather counter-intuitive as we have only $O(1)$ 
numbers in the mass matrix.  One can qualitatively understand this 
point as a consequence of two simple facts.  The first is that one can 
regard $(m_{1},m_{2},m_{3})$ as a randomly oriented vector inside the 
sphere.  Consider only the octant where all components are positive.  
Then it is easy to see that it does not happen very often to have the 
vector pointing toward the center of the octant where all eigenvalues 
are $O(1)$.  It often fluctuates to be close to one of the planes, 
where one of the eigenvalues becomes small compared to others.  But 
having two eigenvalues small requires the vector to point close to one 
of the axes, which is less likely.  The second fact is that the 
measure has the factor 
$(m_{1}^{2}-m_{2}^{2})(m_{2}^{2}-m_{3}^{2})(m_{3}^{2}-m_{1}^{2})$ 
which further pulls three eigenvalues apart from each other.  As 
discussed in more detail in the Appendix, this factor reflects the 
fact that the diagonalization matrix becomes undetermined when two of 
the eigenvalues become degenerate, and therefore it must be there for 
any choice of the measure.  Unless the unknown weight factor which may 
be present on top of the linear measure has singularities when some of 
the eigenvalues coincide, the effect of this factor remains.  
Therefore, obtaining small hierarchy among three eigenvalues is a 
rather general consequence of random matrices.

\begin{figure}
	\centerline{
	\psfig{file=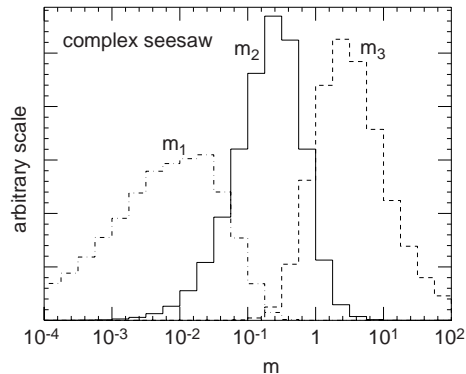,width=0.4\textwidth}
	}
	\caption{The distributions of mass eigenvalues for the complex 
	seesaw case with the linear measure.}
	\label{fig:slog10m}
\end{figure}

The second distribution we show is the mass spectrum in the complex 
seesaw case, with the measure Eq.~(\ref{eq:seesawc}).  Recall that one 
needs to diagonalize the seesaw mass matrix Eq.~(\ref{eq:seesawm}) to 
obtain mass eigenvalues of light (dominantly left-handed) neutrino 
states.  In Fig.~\ref{fig:slog10m}, we show distributions in three 
mass eigenvalues.  The hierarchy among the eigenvalues is larger than 
that in the complex Majorana case.  First of all, there is no cutoff at 
$m_{3}=1$ because the seesaw formula involves the inverse of the mass 
matrix, whose eigenvalue can be larger than 1.  The smaller 
hierarchies in the right-handed Majorana mass matrix and the Dirac 
mass matrix work together to produce a larger hierarchy in the seesaw 
mass matrix.  

\begin{figure}
	\centerline{
	\psfig{file=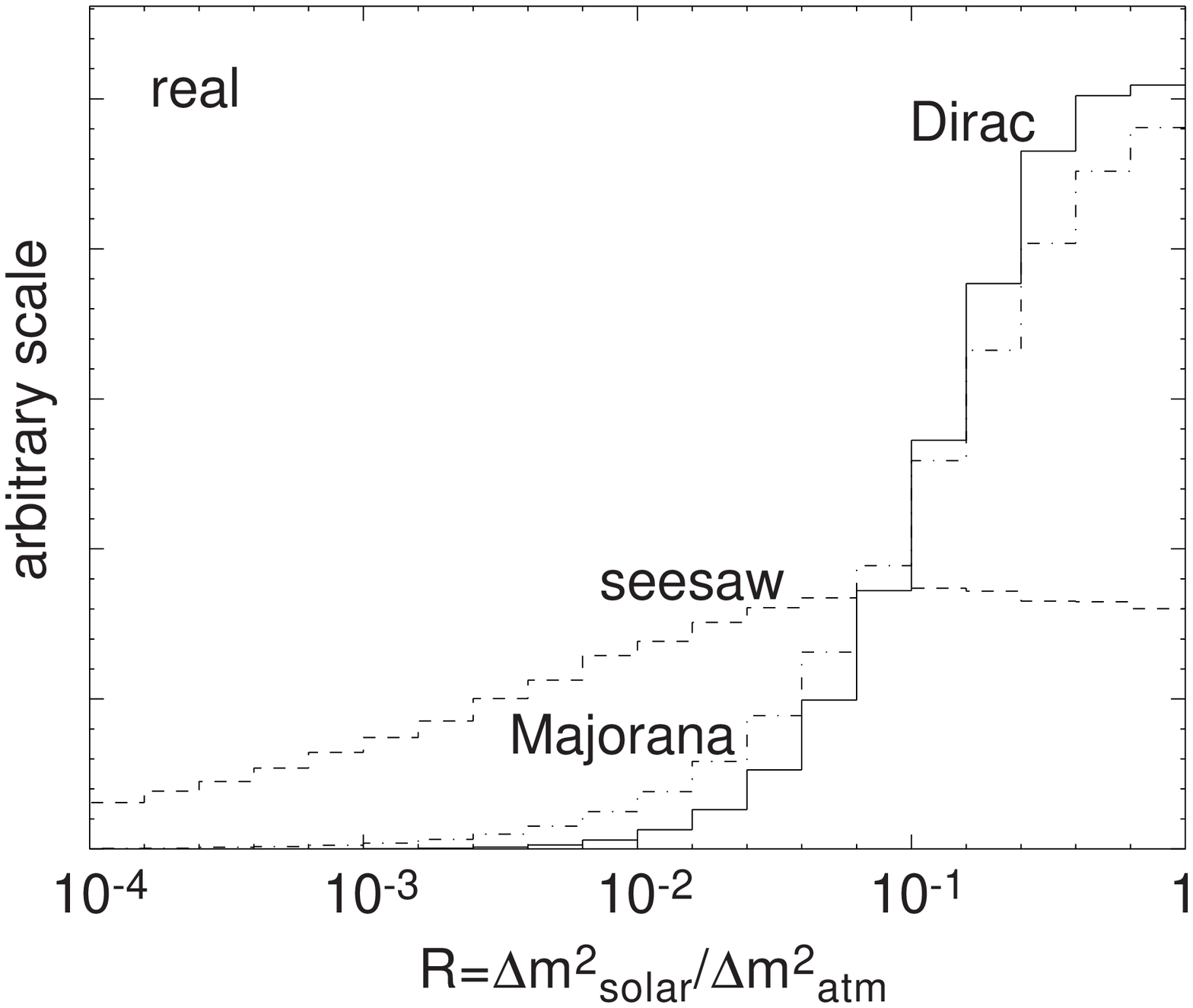,width=0.4\textwidth}\hfill
	\psfig{file=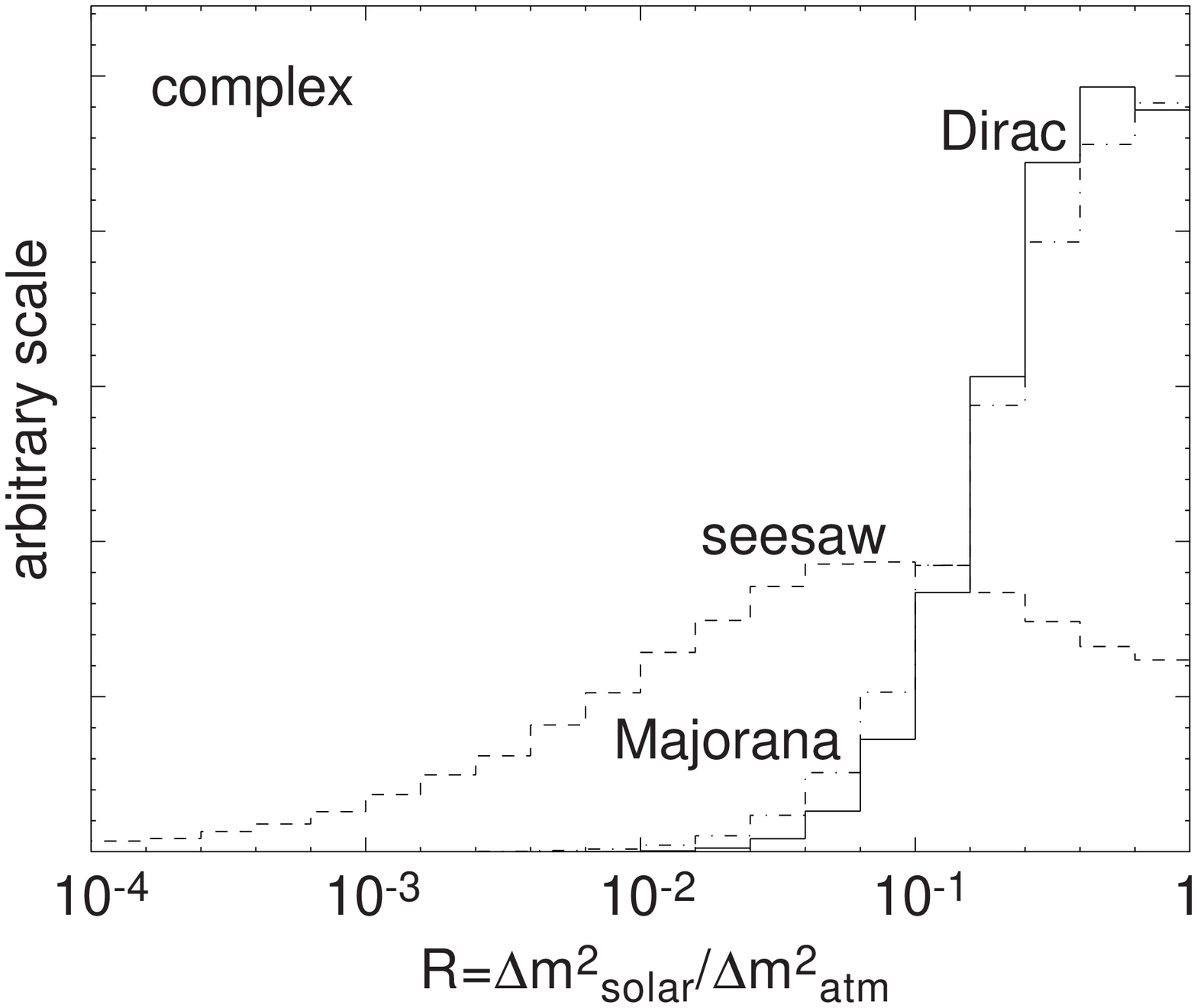,width=0.4\textwidth}
	}
	\caption{The distribution of the ratio $R$ of two $\Delta m^{2}$ for 
	real and complex mass matrices.}
	\label{fig:R}
\end{figure}

Since the anarchy does not specify the overall scale of the neutrino 
mass matrix elements, the useful quantity is the ratio of two $\Delta 
m^{2}$ to compare the consequence of the anarchy and the 
phenomenology of neutrino oscillation.  The Fig.~\ref{fig:R} shows the 
ratio of two $\Delta m^{2}$: $R = 
(m_{2}^{2}-m_{1}^{2})/(m_{3}^{2}-m_{2}^{2})$ for most of the time 
($\sim 90\%$), but when $R>1$, we switch two $\Delta m^{2}$ and define 
$R = (m_{3}^{2}-m_{2}^{2})/(m_{2}^{2}-m_{1}^{2})$.  We will compare 
this ratio to the observed hierarchy between two $\Delta m^{2}$ 
relevant to the solar and atmospheric neutrino oscillations.  For both 
real and complex mass matrices, the Dirac and Majorana cases prefer 
two $\Delta m^{2}$ to be close to each other, while the seesaw cases 
prefer a hierarchy between two $\Delta m^{2}$ of about a factor of 
20.  This behavior could be expected from the distribution of mass 
eigenvalues Figs.~\ref{fig:mlog10m} and \ref{fig:slog10m}.

\section{Neutrino Phenomenology\label{sec:phenomenology}}

Now we are in the position to compare the consequences of the anarchy 
to the phenomenology of neutrino oscillation.

\subsection{Is The Near-Maximal Mixing Special?}

One of the initial reaction to the apparent maximal mixing in the 
atmospheric data is why the nature prefers to be at the boundary of 
the physical region: $0 \leq \sin^2 2\theta \leq 1$.  This observation 
led to the wide-spread notion that the neutrino mass matrix must have 
a very particular structure.  We point out in this section that this 
reaction is based on the use of the incorrect parameter space.  As we 
have seen above, the ``correct'' parameter to use is $\theta$ for real 
matrices or $\sin^{2} \theta$ for complex matrices rather than 
$\sin^{2} 2\theta$.  Once the correct parameter is used, the current 
atmospheric data do not seem to require any particular structure.

First we point out that the apparent maximal mixing is not at the 
boundary of the physical region, but rather at the center of the 
parameter space.  Traditionally, the neutrino oscillation data had 
been plotted against $\sin^{2} 2\theta$, but this covers only a half 
of the parameter space $0 \leq \theta \leq \frac{\pi}{4}$ (the light side).  
There is another half $\frac{\pi}{4} \leq \theta \leq \frac{\pi}{2}$ 
(the dark side) which is physically distinct from the light side as we 
will briefly explain below.

Neutrino oscillations occur if neutrino mass eigenstates are different 
from neutrino weak eigenstates.  Assuming that only two neutrino 
states mix, the relation between mass eigenstates ($\nu_{1}$ and 
$\nu_{2}$) and flavor eigenstates (for example $\nu_{e}$ and 
$\nu_{\mu}$) is simply given by
\begin{eqnarray}
\label{masseigen}
|\nu_1\rangle&=&\cos\theta|\nu_e\rangle-\sin\theta|\nu_{\mu}\rangle,
\nonumber \\
|\nu_2\rangle&=&\sin\theta|\nu_e\rangle+\cos\theta|\nu_{\mu}\rangle,
\end{eqnarray}
where $\theta$ is the vacuum mixing angle.  The mass-squared 
difference is defined as $\Delta m^2\equiv m_2^2-m_1^2$.  We are 
interested in the range of parameters that encompasses all physically 
different situations.  First, observe that Eq.~(\ref{masseigen}) is 
invariant under $\theta \rightarrow \theta+\pi$, $|\nu_e\rangle 
\rightarrow -|\nu_e\rangle$, $|\nu_\mu\rangle \rightarrow 
-|\nu_\mu\rangle$, and hence the ranges 
$[-\frac{\pi}{2},\frac{\pi}{2}]$ and $[\frac{\pi}{2},\frac{3\pi}{2}]$ 
are physically equivalent.  Next, note that it is also invariant under 
$\theta \rightarrow -\theta$, $|\nu_\mu\rangle \rightarrow 
-|\nu_\mu\rangle$, $|\nu_2\rangle \rightarrow -|\nu_2\rangle$, hence 
it is sufficient to only consider $\theta\in [0,\frac{\pi}{2}]$.  
Finally, it can also be made invariant under $\theta \rightarrow 
\frac{\pi}{2} - \theta$, $|\nu_\mu\rangle \rightarrow 
-|\nu_\mu\rangle$ by relabeling the mass eigenstates $|\nu_1\rangle 
\leftrightarrow |\nu_2\rangle$, {\it i.e.} $\Delta m^2 \rightarrow 
-\Delta m^2$.  Thus, we can take ($\Delta m^2>0$) without a loss of 
generality.  All physically different situations are obtained by 
allowing $0\leq \theta\leq \frac{\pi}{2}$.  This very simple yet often 
forgotten point, that maximal mixing $\theta= \pi/4$ is not at the 
boundary of the parameter space but rather at the center of the 
continuum, has important implications on the perception.  This has 
already been emphasized in the context of model building 
\cite{models}.

For the case of oscillations in the vacuum, the survival probability
is given by
\begin{equation}
        P(\nu_{e}\rightarrow \nu_{e}) = 1 - \sin^{2} 2\theta
                \sin^{2} \left( 1.27 \frac{\Delta m^{2}}{E}L \right).
        \label{eq:P}
\end{equation}
Here, $\Delta m^{2}$ is given in eV$^{2}/c^{4}$, $E$ in GeV, and $L$ 
in km.  In this case the oscillation phenomenon can be parameterized 
by $\Delta m^{2}$ and $\sin^{2} 2\theta$, since $\theta$ and 
$\frac{\pi}{2}-\theta$ yield identical survival probabilities.  
Therefore we can restrict ourselves to $0\leq \theta\leq 
\frac{\pi}{4}$, and use the parameter space $(\Delta m^{2}, \sin^{2} 2 
\theta)$ without any ambiguity.  This is indeed an adequate 
parameterization for reactor antineutrino oscillation experiments, 
short-baseline accelerator neutrino oscillation experiments, and 
$\nu_{\mu} \leftrightarrow \nu_{\tau}$ atmospheric neutrino 
oscillation experiments.

On the other hand, the effect of matter in neutrino propagation
clearly distinguishes the light side from the dark side.  One such
case is the so-called MSW (Mikheyev--Smirnov--Wolfenstein) effect on
the neutrino propagation in the Sun.   The survival probability of the
electron-neutrino is given by \cite{darkside}
\begin{eqnarray}
        P(\nu_{e}\rightarrow \nu_{e}) & = & 
        P_{1} \cos^{2} \theta + (1-P_{1}) \sin^{2} \theta \nonumber \\
        & &
        - \sqrt{P_{c} (1-P_{c})} \cos 2\theta_M \sin 2 \theta
        \cos \left( 2.54 \frac{\Delta m^{2}}{E}L + \delta \right),
        \label{eq:Pmatter}
\end{eqnarray}
where $P_c$ is the hopping probability, $\theta_M$ is the mixing angle
at the production point, $P_1 = P_c \sin^2 \theta_M + (1-P_c) \cos^2
\theta_M$, and $\delta$ is a phase induced by the matter effects,
which is not important for our purposes.  See Ref.~\cite{seasonal,us}
for notation.  One can write down an analytic formula for the
electron-neutrino survival probability if the electron number density
profile in the Sun is approximated by an exponential $n_e \propto
e^{-r/r_0}$:
\begin{equation}
  P_c = \frac{e^{-\gamma \sin^2 \theta} - e^{-\gamma}}{1-e^{-\gamma}},
\end{equation}
where $\gamma = 2\pi r_0 \Delta m^2/2E$.  It is easy to check that the 
survival probability is invariant under the simultaneous change 
$\Delta m^2 \rightarrow -\Delta m^2$ and $\theta \rightarrow 
\frac{\pi}{2} - \theta$, but not individually.  Therefore the matter 
effect distinguishes the light and the dark sides.

It cannot be overemphasized that physics is different between the
light and the dark side even if the neutrino propagates in the vacuum.
It is just that the vacuum oscillation probability comes out the same
in both sides.  In the case of $\nu_\mu$--$\nu_\tau$ mixing, for
instance, one can ask the question if the mass eigenstate closer to
the $\nu_\mu$ state is heavier or lighter.  If one can literally weigh
them, this question has a clear physical meaning.  The fact that the
vacuum oscillation does not distinguish both sides is simply due to
the focus on one particular physical quantity.

\subsection{Atmospheric Neutrino Data}

\begin{figure}
	\centerline{
	\psfig{file=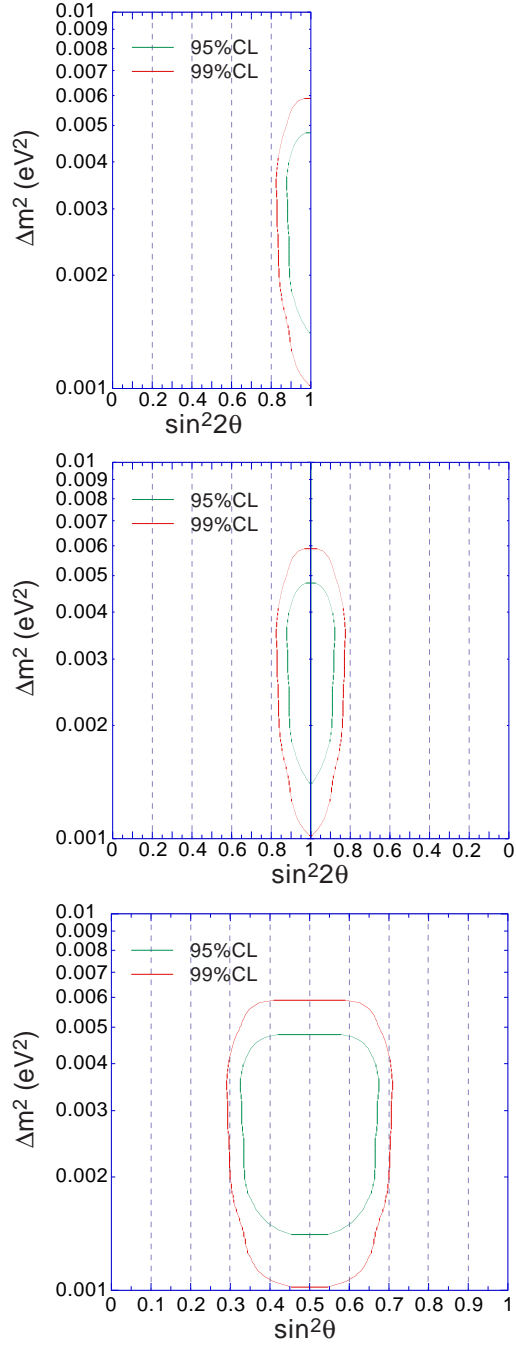,width=0.45\textwidth}
	}
	\caption{The preferred region from the fit to the atmospheric 
	neutrino data from the SuperKamiokande experiment at 90\% and 99\% 
	C.L. (a) As it is presented by the collaboration.  (b) The dark 
	side is added.  (c) Replotted on $\sin^{2} \theta$.  The angle 
	$\theta$ is approximately $\theta_{23}$ if $\theta_{13}$ is small 
	enough.}
	\label{fig:atmos}
\end{figure}

Given the fact that the light and the dark sides are physically 
distinct, it is useful to plot the preferred parameter range from the 
SuperKamiokande experiment continuously from $0^\circ$ to $90^\circ$.  
In Fig.~\ref{fig:atmos}, we show plots on $\sin^2 2\theta$ and 
$\sin^2 \theta$ axes.  Even though the preferred region appears very 
close to the maximal angle with the $\sin^2 2\theta$ plot, it does not 
on the other plots.  In fact, we argued in the previous section that 
the plot should be shown on the $\sin^{2}\theta$ (rather than 
$\sin^{2} 2\theta$) parameter.

The plots clearly show that the region preferred by the 
SuperKamiokande atmospheric neutrino data is a somewhat large ``blob'' 
at the center of the parameter space, and is not necessarily special.  
At this point, we can say that a random mass matrix can well produce 
the apparent ``near-maximal'' angle.  In this sense, we have been 
fooled by the plots on $\sin^{2} 2\theta$ because it shrinks the large 
angle region.

The atmospheric neutrino data sets the scale for the larger of two 
$\Delta m^{2}$ to be at 1--6$\times 10^{-3}$~eV$^{2}$.

\subsection{Solar Neutrino Data}

\begin{figure}
	\centerline{
	\psfig{file=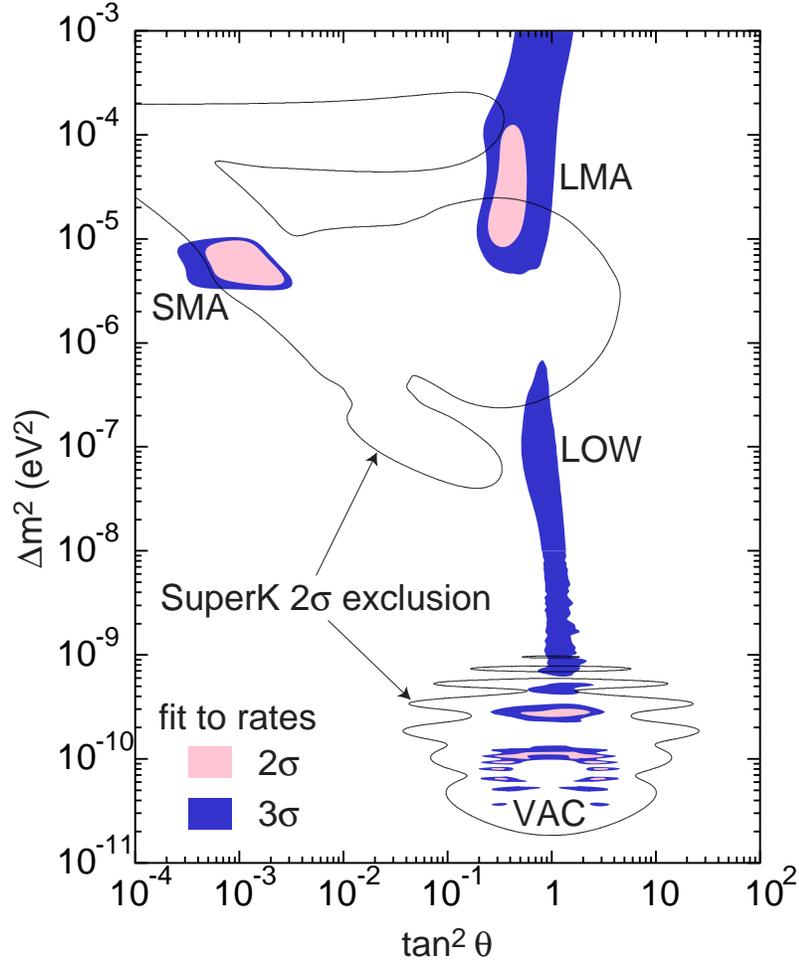,width=0.8\textwidth}
	}
	\caption{The preferred regions from the fit to the solar 
	neutrino rates SuperKamiokande, Gallium experiments, and Homestake, 
	taken from Ref.~\cite{darkside}.  Overlaid is the 95\% CL exclusion 
	from the SuperKamiokande day/night energy spectra \cite{Koshio}.  The 
	angle $\theta$ is approximately $\theta_{12}$ if $\theta_{13}$ is 
	small enough.}
	\label{fig:solar}
\end{figure}

As for the solar neutrino problem, the small hierarchy among $\Delta
m^{2}$ in the seesaw case (see Fig.~\ref{fig:R}) strongly favor the
Large Mixing Angle (LMA) MSW solution which requires $\Delta m^{2} =
10^{-5}$--$10^{-3}$~eV$^{2}$, especially its upper end
(Fig.~\ref{fig:solar}).  Recall that $\Delta m^{2}$ for the
atmospheric neutrino oscillation is at $10^{-3}$--$10^{-2}$~eV$^{2}$ as
discussed in the previous section (Fig.~\ref{fig:atmos}).  Since we
know that $\theta_{13}$ has to be somewhat small (we will come back to
this in the next subsection), two-flavor analysis of the solar
neutrino oscillation is basically valid even when three flavors are
considered.  For the LMA solution, the mixing angle is $\sin^{2}
2\theta = 0.5$--$1$.  Recent analysis including the dark side suggest
$\tan^{2} \theta = 0.15$--$2$ \cite{concha}.  This is the region where
the distribution in $\theta_{12}$ is indeed peaked.  Therefore, it is
fair to say that the LMA solution is ``right on the mark'' for the
anarchy with no physical distinction among three generations of
left-handed lepton doublets.

\subsection{Reactor Neutrino Data}

\begin{figure}
	\centerline{
	\psfig{file=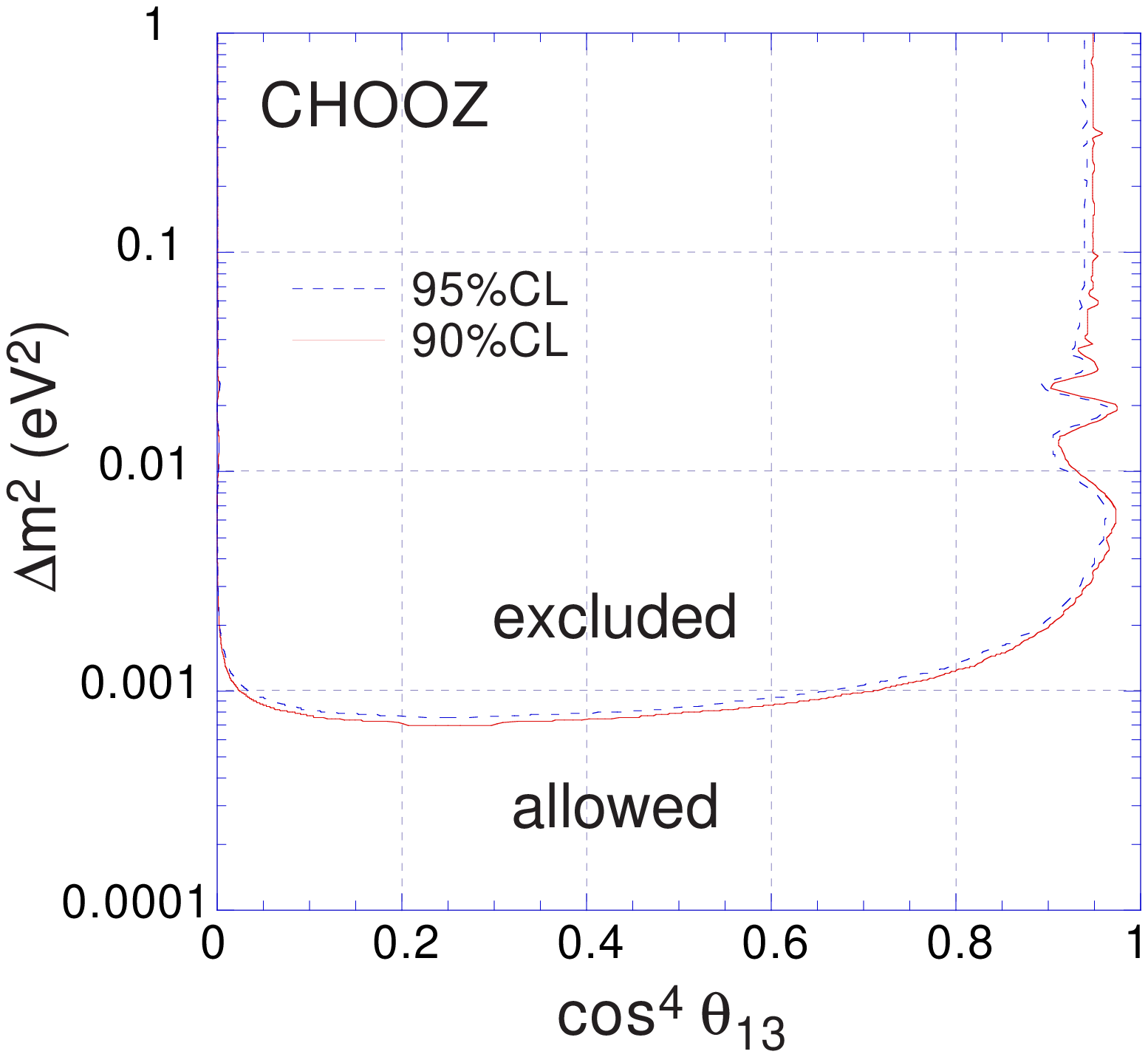,width=0.6\textwidth}
	}
	\caption{The limit on $\theta_{13}$ from the CHOOZ reactor 
	anti-neutrino oscillation experiment.  It is plotted against 
	$\cos^{4} \theta_{13}$.}
	\label{fig:CHOOZ}
\end{figure}

The only quantity where the distribution tends to go against the 
experimental data is the angle $\theta_{13}$.  The lack of 
$\bar{\nu}_{e}$ disappearance in the reactor neutrino experiments such 
as CHOOZ and Palo Verde had placed a limit on 
$|U_{e3}|=\sin\theta_{13}$ (in case $\Delta m^{2}_{12}>\Delta 
m^{2}_{23}$, it is on $|U_{e1}|$).  In Fig.~\ref{fig:CHOOZ}, we replot 
the constraint from the CHOOZ experiment \cite{CHOOZ} against 
$\cos^{4} \theta_{13}$ which is the variable that makes the 
distribution flat according to the $U(3)$ Haar measure.  Depending on 
the precise value of $\Delta m^{2}$ for the atmospheric neutrino 
oscillation, one obtains $\cos^{4} \theta_{13} > 0.7$--$0.95$.  The 
experimentally allowed window is hence 5--30\%.  We, however, do not see 
this as a fatal problem for the anarchy.  Three out of four physical 
quantities, $\sin^{2} 2\theta_{23}$, $\sin^{2} 2\theta_{12}$, and 
$(\Delta m^{2})_{\odot}/(\Delta m^{2})_{\rm atm}$ worked out right on, 
and the last one needs a little bit of fluctuation.  What this means 
instead is that $U_{e3}$ may well be just below the current limit.

\section{Hierarchy\label{sec:hierarchy}}

In the previous subsections, we have shown that we expect large mixing 
angles, quite often ``maximal,'' and only a small hierarchy among mass 
eigenvalues when there is no fundamental distinction among three 
generations.  The next obvious question then is what generates small 
angles and large hierarchy for quarks.  The charged leptons also have 
a large hierarchy in their mass eigenvalues.

The most promising approach to this questions is probably an
approximate flavor symmetry.  The idea is that there is a new symmetry
(and hence a new conservation law) which forbids all Yukawa couplings
but for the top quark.  The symmetry is assumed to be broken by a
small parameter, and other Yukawa couplings are generated only at
certain powers of the small symmetry breaking parameter.  As
emphasized in the Introduction, the top Yukawa coupling (and possibly
bottom and tau Yukawa couplings if $\tan\beta \sim 60$ in two-doublet
Higgs models including the MSSM) is regarded as of natural size
$O(1)$, and all other Yukawa couplings are small and hence require
explanations.  The virtue of the approximate flavor symmetry is that
it explains the hierarchical mass eigenvalues as well as the small
mixings at the same time.

Then our view of the fermion masses and mixings are simple: there is
no distinction among three left-handed lepton doublets $L_i$, while
there is for right-handed lepton singlets $e_i$ as well as quarks.
The lack of distinction among $L_i$ would appear ``anarchical''
leading to large mixing angles and small hierarchies in the neutrino
sector.  However the distinction among three $e_i$ as well as quarks
due to their different flavor quantum numbers lead to hierarchical
masses and small quark mixing angles.  This was the simple model
presented in \cite{anarchy}.

The most simple extension of such model to the whole fermion masses is
based on the following $U(1)$ flavor symmetry.  The charge
assignment is $SU(5)$-like:\footnote{This charge assignments would
  prefer a large $\tan\beta$.  Another possibility is to assign charge
  ${\bf +1}$ for all $\underline{5}$'s which would prefer a small
  $\tan\beta$.}
\begin{equation}
  \begin{array}{ccc}
    \underline{10}_1 ({\bf +2}) & \underline{10}_2 ({\bf +1}) &
    \underline{10}_3 ({\bf 0}) \\
    \underline{5}^*_1 ({\bf 0}) & \underline{5}^*_2 ({\bf 0}) &
    \underline{5}^*_3 ({\bf 0}) \\
    \underline{1}_1 ({\bf 0}) & \underline{1}_2 ({\bf 0}) &
    \underline{1}_3 ({\bf 0})
  \end{array}
\end{equation}
where the subscripts are generation indices and the $U(1)$ flavor
charges are given in bold face.  The $SU(5)$-like multiplets contain
$\underline{10} = (Q, u^c, e^c)$, $\underline{5}^* = (L, d^c)$, and
$\underline{1} = N^c$ where $N=\nu_R$ is the right-handed neutrino
state.  The idea here is that the hierarchy in the fermion masses and
mixings are solely due to the $\underline{10}$'s.  Assuming that the
$U(1)$ flavor symmetry is broken by a single parameter $\epsilon ({\bf
  -1}) \sim 0.04 \sim \lambda^2$, the Yukawa matrices read
\begin{equation}
  Y_u \sim \left( \begin{array}{ccc} 
      \epsilon^4 & \epsilon^3 & \epsilon^2\\
      \epsilon^3 & \epsilon^2 & \epsilon\\
      \epsilon^2 & \epsilon & 1
      \end{array} \right), \,
  Y_d \sim \left( \begin{array}{ccc} 
      \epsilon^2 & \epsilon^2 & \epsilon^2\\
      \epsilon & \epsilon & \epsilon\\
      1 & 1 & 1
      \end{array} \right), \,
  Y_l \sim \left( \begin{array}{ccc} 
      \epsilon^2 & \epsilon & 1\\
      \epsilon^2 & \epsilon & 1\\
      \epsilon^2 & \epsilon & 1
      \end{array} \right), \,
  Y_\nu \sim \left( \begin{array}{ccc} 
      1 & 1 & 1\\
      1 & 1 & 1\\
      1 & 1 & 1
      \end{array} \right), 
\end{equation}
where the left-handed (right-handed) fields couple to them from the
left (right) of the matrices.  There are ``random'' $O(1)$
coefficients in each of the matrix elements.  The property that $Y_d
\sim Y_l^T$ is true in many $SU(5)$-like models.  Finally, the
Majorana mass matrix of right-handed neutrinos is
\begin{equation}
  M_R \sim M_0 \left( \begin{array}{ccc} 
      1 & 1 & 1\\
      1 & 1 & 1\\
      1 & 1 & 1
      \end{array} \right), 
\end{equation}
where $M_0 \sim 10^{15}$~GeV is the mass scale of lepton-number
violation.  

Interestingly, these mass matrices had been already discussed in the
literature \cite{BB}, but there has been a perception that the $O(1)$
coefficients as well as texture zeros have to be chosen very carefully.
Similar mass matrices had been obtained in the context of composite
models \cite{Strassler,Haba}, extra dimensions \cite{Yoshioka}, and
anomalous $U(1)$ \cite{HKN}.  We certainly do not claim that the above
mass matrices are new.  However, before the analysis in
\cite{anarchy}, these mass matrices had not been taken seriously as
they were not believed to explain the atmospheric and solar neutrino
oscillations simultaneously unless careful adjustments and/or more
flavor symmetries are imposed in the above-cited papers.  Our proposal
is that this simple flavor $U(1)$ symmetry is enough to understand the
observed pattern of quark, lepton masses, quark mixings and neutrino
oscillations based on the simple assumption that the $O(1)$
coefficients would appear pseudo-random from the low-energy point of
view.

The above mass matrices would naturally explain (1) the ``double''
hierarchy in up quarks relative to the hierarchy in down quarks and
charged leptons, (2) $V_{cb} \sim O(\epsilon) \sim O(\lambda^2)$, (3)
the similarity between the down quark and charged lepton masses.  Some
``concerns'' with the above mass matrices would be that the following
points may be difficult to understand: (a) $m_s \sim m_\mu/3$, (b)
$m_e \sim m_d/3$, (c) $V_{us} \sim \epsilon^{1/2}$ rather than
$\epsilon$.  However, in view of the fact that the $O(1)$ coefficients
would seem ``anarchical'' from the low-energy point of view, a factor
of $1/3$ is quite likely to appear.  And once $m_s$ is fluctuated
downwards by a factor of $\sim 1/3$, $V_{us}$ would fluctuate upwards
to $\sim 3\epsilon$ which is enough to understand the observed pattern
of masses and mixings.

What if the $SU(5)$-GUT is true?  One can still understand the
pseudo-randomness of the $O(1)$ coefficients in the down and lepton
masses in the following fashion.  Suppose there are many vector-like
multiplets at the GUT-scale which do not survive below the GUT-scale
simply because of their vector-like nature (``survival hypothesis''
\cite{survival}).  However, they can mix significantly with our quarks and
leptons as long as they share the same flavor quantum numbers.  The
mixing is likely to pick up the $SU(5)$-breaking order parameter, such
as the vacuum expectation value of the $SU(5)$-adjoint Higgs boson.
Then the resulting low-energy Yukawa couplings would not respect
$SU(5)$ invariance and hence would appear random independently between
the down quarks and charged leptons.  It is simply that the complexity
of physics at the GUT-scale leads to a pseudo-random nature of the
$O(1)$ coefficients from the low-energy point of view.

Note that we could have assigned non-trivial charges to right-handed
neutrino states, and would obtain exactly the same phenomenologies for
quarks, charged leptons and neutrino oscillations.  

\section{Conclusions\label{sec:conclusions}}

We have advocated a new approach to build models of fermion masses and 
mixings, namely anarchy.  The approach relies only on the approximate 
flavor symmetries, and scan the $O(1)$ coefficients randomly.  The 
randomness in $O(1)$ coupling constants is indeed what one expects 
in models which are sufficiently complicated or which have a large 
number of fields mixed with each other.

This approach is particularly useful for neutrinos.  Assuming there is 
no physical distinction among three generations of neutrinos, we have 
shown that the probability distributions in MNS mixing angles could be 
predicted independent of the choice of the measure.  This is because 
the mixing angles are distributed to the Haar measure of the Lie 
groups whose elements diagonalize the mass matrices.  The near-maximal 
mixings, as observed in the atmospheric neutrino data and is required 
in the LMA solution to the solar neutrino problem, are highly 
probable.  Even though somewhat more dependent on the particular 
choice of the measure, the distributions in the mass spectrum can also 
be worked out.  A small hierarchy between the $\Delta m^{2}$ for the 
atmospheric and the solar neutrinos is obtained very easily; the 
complex seesaw case gives a hierarchy of a factor of 20 as the most 
probable one.  Therefore, these three observables are nicely 
consistent with the current experimental data.  On the other hand, the 
$U_{e3}$ angle must lie in the 10\% tail of the distribution.  It is 
not surprising, however, that three out of four observables are 
``right on the mark,'' while the fourth one is off at the 10\% level.  
In other words, we expect $U_{e3}$ to be just below the current limit 
from the CHOOZ experiment.  This is an ideal situation for the future 
long-baseline neutrino oscillation experiments.  Moreover, 
CP-violating parameter $\sin \delta$ is preferred to be maximal.

The anarchy can be easily extended to the charged lepton and quark 
sectors.  We presented a simple $SU(5)$-like flavor charge assignment 
which works well phenomenologically.  

\appendix

\section{Measures\label{sec:measures}}

\subsection{Decomposition of the Measures\label{subsec:decomposition}}

The measure over the mass matrix elements can be studied analytically.
We consider Majorana and Dirac cases with both real and complex matrix
elements.  The aim here is to decompose the measure into that over the 
diagonalization matrix and other over the mass eigenvalues.

First of all, it is important to note that the linear measure over the
mass matrix elements is invariant under change of basis.  This is
because a $N \times N$ real Majorana mass matrix transforms as a
symmetric rank-two tensor under the $O(N)$ rotation of the basis $M
\rightarrow O M O^T$, and hence a linear representation.  The naive
$N(N+1)/2$-dimensional measure $dM$ is hence invariant under this
$O(N)$ rotation.  Similarly, a complex Majorana mass matrix transforms
as a symmetric rank-two tensor under the $U(N)$ rotation $M
\rightarrow U M U^T$, a real Dirac mass matrix as
$(\underline{N},\underline{N})$ under $O(N) \times O(N)$ as $M
\rightarrow O_L M O_R^T$, and a complex Dirac mass matrix as
$(\underline{N},\underline{N})$ under $U(N) \times U(N)$ as $M
\rightarrow U_L M U_R^T$.  

This observation allows us to separate the measure over the mass matrix
elements in terms of eigenvalues and group transformations.  Starting
with the real Majorana case, any mass matrix can be written as
\begin{equation}
  M = O D O^T,
\end{equation}
where $O \in SO(N)$ and $D = \mbox{diag}(m_1, m_2, \cdots, m_N)$ is a
real diagonal matrix.  Because of the invariance of the measure $dM$
under the change of basis $M \rightarrow O_1 M O_1^T$, the measure
$dM$ should contain the measure over the group $dO$ which is invariant
under the left transformation $O \rightarrow O_1 O$.  Since $SO(N)$ is
a compact Lie group, a left-invariant measure is also right-invariant.
Therefore, $dO$ should be the invariant Haar measure over the group
$SO(N)$.  Then the measure $dM$ can be written as
\begin{equation}
  dM = f(m_1, \cdots, m_N) \prod_{i=1}^N dm_i dO.
\end{equation}
Here the yet-undetermined function $f$ is symmetric under the
interchange of eigenvalues.  However, when two of the eigenvalues
coincide, an $SO(2)$ subgroup of the $SO(N)$ becomes ill-defined and
we expect zeros in the function $f$.  Therefore, $f \propto
\prod_{i<j}^N (m_i-m_j)$.  This factor already saturates the dimension
of the measure, and hence
\begin{equation}
  dM = \prod_{i<j} (m_i-m_j) \prod_{i=1}^N dm_i dO
\end{equation}
up to a constant normalization factor.  As we will see below, this
simple expression allows us to study the distributions in the mixing
angles and eigenvalues very easily.

A similar consideration for the real Dirac case leads to the
following measure:
\begin{equation}
  dM = \prod_{i<j} (m_i^2-m_j^2) \prod_{i=1}^N dm_i dO_L dO_R.
\end{equation}
Here, the mass matrix is parameterized as $M = O_L D O_R^T$, and
$dO_{L,R}$ are Haar measures of both $O(N)$ factors. If we restrict
$O_{L,R}$ matrices to $SO(N)$, we must allow both positive and
negative eigenvalues.  However, in the vicinity of $O_{L,R}=1$, one
can change the basis by an $O(N)$ rotation $\mbox{diag}(1, \cdots, 1,
-1, 1, \cdots, 1)$ to flip sign of one of the eigenvalues, and hence
$m_i = -m_j$ should give a singularity as well as $m_i = m_j$.
Therefore, the prefactor should be $\prod_{i<j} (m_i^2-m_j^2)$ which
saturates the dimension and hence is unique up to a normalization
constant.

The complex Majorana case is given by
\begin{equation}
  dM = \prod_{i<j} (m_i^2-m_j^2) \prod_{i=1}^N m_i dm_i dU.
\end{equation}
The mass matrix is parameterized as $M = U D U^T$ where $U \in U(N)$.
The prefactor can be determined by the same argument as in the real
Dirac case, because the sign of an eigenvalue can be flipped by an
$U(N)$ rotation ${\rm diag}(1, \cdots, 1, i, 1, \cdots, 1)$.  However
this does not saturate the dimension of the measure, leaving
additional $N$ powers of the eigenvalues.  Only such an additional
factor which is symmetric under the interchange and also invariant
under the change of the sign of an eigenvalue up to an overall sign is
the product of all eigenvalues, $\prod=i^N m_i$.  Therefore the above
parameterization is unique.

Finally the complex Dirac case gives
\begin{equation}
  dM = \prod_{i<j} (m_i^2-m_j^2)^2 \prod_{i=1}^N m_i dm_i \frac{dU_L
  dU_R}{\prod_i^N d\phi_i}.
\end{equation}
The parameterization is $M = U_L D U_R^\dagger$.  The prefactor is 
based on the same argument as the real Dirac and complex Majorana 
cases, and the reason why $(m_i^2-m_j^2)$ factor is squared is because 
$U(2)$ subgroups of $U_{L}$ and $U_{R}$ rotations are not independent 
when $m_{i}=m_{j}$.  We verified this for $N=2$, and this conjecture 
appears to be the unique generalization to arbitrary $N$.  The group 
measure is modded out by the following transformation $U_{L,R} 
\rightarrow T U_{L,R}$, where $T = \mbox{diag}(e^{i\phi_1}, 
e^{i\phi_2}, \cdots, e^{i\phi_N})$, because this simultaneous change 
of $U_L$ and $U_R$ does not change the mass matrix $M$.

Note that we do not quite know on what measure we should regard the
mass matrix elements to be ``random.''  The measure we discussed
above, namely the linear measure over the mass matrix elements may not
necessarily be the correct measure.  However, the requirement of the
basis independence implies that the invariant Haar measure is
certainly a part of it.  The only possible modification of the linear
measure discussed above is a weight function which depends on the
symmetric polynomials of the eigenvalues.  Therefore, the distributions
on the mass eigenvalues can change from the linear measure to the
``correct'' one, but the distributions on the mixing angles cannot.  

\subsection{Haar Measures\label{subsec:Haar}}

The invariant Haar measures of $SO(3)$ and $U(3)$ groups relevant to 
our study can be obtained once their parameterization is fixed.

First define the (right) Maurer--Cartan forms on the group manifold 
$G$
\begin{equation}
	\omega^{a} = -i {\rm Tr} T^{a} U^{-1} d U
\end{equation}
where $U \in G$.  Here, $T^{a}$ are hermitian generators.  It is easy 
to check that these forms are trivially invariant under the 
left-translation of the group $U \rightarrow U_{0} U$.  Under the 
right-translations $U \rightarrow U U_{0}$, however, they transform 
among each other:
\begin{equation}
	\omega^{a} \rightarrow O_{0}^{ab} \omega^{b},
\end{equation}
where
\begin{equation}
	U_{0} T^{a} U_{0}^{-1} = O^{ab} T^{b}.
\end{equation}
Note that $O_{0}$ is an orthogonal matrix, which is easily checked by 
doing a successive left-translation by $U_{0}^{-1}$.  Because the 
Maurer--Cartan forms transform homogeneously by an orthogonal 
matrix, the following combination
\begin{equation}
	dU = \epsilon_{a_{1} a_{2} \cdots a_{N}}
	\omega^{a_{1}}\wedge\omega^{a_{2}}\wedge \cdots \wedge\omega^{a_{N}}
\end{equation}
is invariant under the right-translations.  Here, $\epsilon_{a_{1} 
a_{2} \cdots a_{N}}$ is the totally anti-symmetric tensor.  This 
measure is invariant because of the orthogonal nature of the 
transformation, and $N = {\rm dim}G$.

For $SO(3)$, take the conventional parameterization
\begin{eqnarray}
  O &=& \left( \begin{array}{ccc}
  	1 & 0 & 0\\
	0 & c_{23} & s_{23}\\
	0 & -s_{23} & c_{23}
  \end{array} \right)
  \left( \begin{array}{ccc}
  	c_{13} & 0 & s_{13}\\
	0 & 1 & 0\\
	-s_{13} & 0 & c_{13}
  \end{array}\right)
  \left( \begin{array}{ccc}
  	c_{12} & s_{12} & 0\\
	-s_{12} & c_{12} & 0\\
	0 & 0 & 1 \end{array} \right) \nonumber \\
  &=& \left( \begin{array}{ccc} c_{12} 
    c_{13} & s_{12} c_{13} & s_{13} \\
	-s_{12} c_{23} - c_{12} s_{23} s_{13} & 
	c_{12} c_{23} - s_{12} s_{23} s_{13} & s_{23} c_{13} \\
	s_{12} s_{23} - c_{12} c_{23} s_{13} &
	-c_{12} s_{23} - s_{12} c_{23} s_{13} & c_{23} c_{13}
  \end{array} \right),
\end{eqnarray}
where $c_{12} = \cos \theta_{12}$ {\it etc}\/.  Then the Haar measure is given by
\begin{equation}
  dO = \cos\theta_{13} d\theta_{12} d\theta_{13} d\theta_{23}
\end{equation}
up to an overall normalization factor.  Note that the measure is flat 
in $\theta_{12}$ and $\theta_{23}$ similarly to the two-generation 
Majorana case discussed earlier.  Therefore these distributions are 
peaked in $\sin^2 2\theta_{12}$ and $\sin^2 2\theta_{23}$ both at $0$ 
and $1$.  The measure is different for $\theta_{13}$ being in the 
middle among three rotations.  The reason why the measure is 
degenerate when $\theta_{13} \rightarrow \pi/2$ is that $\theta_{12}$ 
and $\theta_{23}$ are not independent at this point.

For $U(3)$, take a similar parameterization
\begin{equation}
  U = 
  e^{i\eta} e^{i\phi_{1} \lambda_{3} + i \phi_{2} \lambda_{8}}
  \left( \begin{array}{ccc}
  	1 & 0 & 0\\
	0 & c_{23} & s_{23}\\
	0 & -s_{23} & c_{23}
  \end{array} \right)
  \left( \begin{array}{ccc}
  	c_{13} & 0 & s_{13} e^{-i\delta}\\
	0 & 1 & 0\\
	-s_{13}e^{i\delta} & 0 & c_{13}
  \end{array}\right)
  \left( \begin{array}{ccc}
  	c_{12} & s_{12} & 0\\
	-s_{12} & c_{12} & 0\\
	0 & 0 & 1 \end{array} \right)
  e^{i\chi_{1} \lambda_{3} + i \chi_{2} \lambda_{8}},
\end{equation}
where $\lambda_{3} = {\rm diag}(1, -1, 0)$ and $\lambda_{8} = {\rm 
diag} (1, 1, -2)/\sqrt{3}$ are Gell-Mann matrices.  This 
parameterization is designed to make the angles $\eta$, 
$\phi_{1,2}$, $\chi_{1,2}$ unphysical except for the Majorana phases 
in possible lepton-number violating processes.  Then the Haar measure 
is given by
\begin{equation}
	d U = d s_{12}^{2} d c_{13}^{4} d s_{23}^{2} d\delta
	d\eta d\phi_{1} d\phi_{2} d\chi_{1} d\chi_{2}
\end{equation}
up to an overall normalization factor.  For the Dirac mass matrices, 
we need to mod out linear combinations of $\eta$, $\chi_{1}$, 
$\chi_{2}$ from $U_{L}$ and $U_{R}$, but this does not affect the 
measure for physical angles, $\theta_{12}$, $\theta_{13}$, 
$\theta_{23}$, $\delta$.

\section*{Acknowledgments}
 HM thanks Lawrence Hall and Neal Weiner, and both us thank Masako
 Bando, Taichiro Kugo, Jiro Sato for useful discussions.  Yukawa
 Institute for Theoretical Physics, Nagoya University, and Post-Summer
 Institute in Fuji-Yoshida where part of this research was conducted. 
 This work was supported in part by the Department of Energy under
 contract DE--AC03--76SF00098, in part by the National Science
 Foundation under grant PHY-95-14797, in part by the Grant-in-Aid for
 Science Research, Ministry of Education, Science and Culture, Japan
 (No.  12740146, No.  12014208).

\end{document}